\PassOptionsToPackage{svgnames}{xcolor}

\documentclass[preprint,journal]{vgtc}            % preprint (journal style)

\ieeedoi{10.1109/TVCG.2023.3327398}

\onlineid{0}

\vgtccategory{Research}

\vgtcpapertype{please specify}

\title{Design Patterns for Situated Visualization in Augmented Reality}

\author{%
  \authororcid{Benjamin Lee}{0000-0002-1171-4741},
  \authororcid{Michael Sedlmair}{0000-0001-7048-9292}, and
  \authororcid{Dieter Schmalstieg}{0000-0003-2813-2235}
}
\authorfooter{
  \item
  	Benjamin Lee and Michael Sedlmair are with University of Stuttgart.
  	E-mail: {first-name.last-name}@visus.uni-stuttgart.de
  \item
  	Dieter Schmalstieg is with Graz University of Technology and University of Stuttgart.
  	E-mail: schmalstieg@tugraz.at
}

\abstract{Situated visualization has become an increasingly popular research area in the visualization community, fueled by advancements in augmented reality (AR) technology and immersive analytics. Visualizing data in spatial proximity to their physical referents affords new design opportunities and considerations not present in traditional visualization, which researchers are now beginning to explore. However, the AR research community has an extensive history of designing graphics that are displayed in highly physical contexts. In this work, we leverage the richness of AR research and apply it to situated visualization. We derive design patterns which summarize common approaches of visualizing data in situ. The design patterns are based on a survey of 293 papers published in the AR and visualization communities, as well as our own expertise. We discuss design dimensions that help to describe both our patterns and previous work in the literature. This discussion is accompanied by several guidelines which explain how to apply the patterns given the constraints imposed by the real world. We conclude by discussing future research directions that will help establish a complete understanding of the design of situated visualization, including the role of interactivity, tasks, and workflows.
}


\graphicspath{{figs/}{figures/}{pictures/}{images/}{./}} % where to search for the images

\usepackage{tabu}                      % only used for the table example
\usepackage{booktabs}                  % only used for the table example
\usepackage{lipsum}                    % used to generate placeholder text
\usepackage{mwe}                       % used to generate placeholder figures
\usepackage{mathptmx}                  % use matching math font
\usepackage{color}
\usepackage{wrapfig}

\definecolor{mscolor}{rgb}{0,0,0.}

\definecolor{dscolor}{rgb}{0.5,0,0.9}

\teaser{
   \centering
   \includegraphics[width=\linewidth]{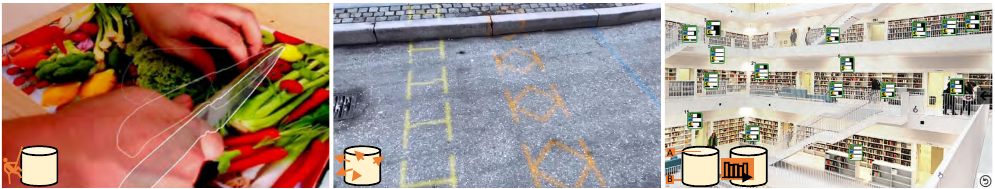}
   \caption{Examples of situated visualization patterns (pattern names in italics):
   (left) motion of the knife indicated using a white outline (\textit{ghost});
   (middle) simulated paint at a construction site \cite{hansenAugmentedRealitySubsurface2021} (\textit{glyphs});
   (right) statistics of sections in a library \cite{tatzgernAdaptiveInformationDensity2016} (\textit{label}, \textit{panel}).
   }
   \label{fig:teaser}
 }

\begin{document}

% -----------------------------------------------------------------------------------------
\firstsection{Introduction}
% -----------------------------------------------------------------------------------------
\maketitle

It has become increasingly common to use immersive displays to visualize information in the space around users---a technique known as \textit{immersive analytics}~\cite{marriottImmersiveAnalytics2018}. A large body of research in immersive analytics has used virtual reality (VR). VR provides a controlled, artificial environment which lets investigators focus on the data, the immersive visual representations, and the interactions they afford. 

In contrast to VR, augmented reality (AR) takes place in the physical world, thus AR may need to consider environmental factors~\cite{satkowskiInvestigatingImpactRealWorld2021,ericksonReviewVisualPerception2020}. For example, a visualization may be automatically positioned to avoid occluding the physical environment~\cite{evangelistabeloAUITAdaptiveUser2022}. The physical environment is not just a nuisance; it can, in fact, be a major asset. AR commonly involves a semantic relationship~\cite{whiteInteractionPresentationTechniques2009} between visualizations and physical \textit{referents}---which are the places and entities to which data corresponds~\cite{willettEmbeddedDataRepresentations2017}.
This is referred to as \textit{situated visualization}~\cite{whiteSiteLensSituatedVisualization2009, whiteInteractionPresentationTechniques2009}.
Depending on the spatial proximity between visualization and referent, Willett et al.~\cite{willettEmbeddedDataRepresentations2017} classify visualizations into non-situated (i.e.,~in different locations), situated (i.e.,~in the same location), and embedded (i.e.,~directly on top or adjacent to the referent). Situated and embedded visualizations allow users to access, view, and understand data in situ, without splitting attention across physical and virtual objects.

The choice of how to position a visualization with respect to its referent matters. Willett et al.~\cite{willettEmbeddedDataRepresentations2017} describe varying levels of indirection, i.e.,~the (perceived) distance between the visualization and the referent. Therefore, a situated visualization designer needs to decide on an appropriate level of indirection to ensure perceptibility and usability of both the visualization and the referent, while also maintaining the semantic relationship via spatial proximity.

%In contrast to the growing field of (immersive) situated visualization, %%% potentially controversial
AR researchers have been exploring ways of displaying information in physical contexts for decades, such as in physical assembly (e.g.,~\cite{lochComparingVideoAugmented2016,stanescuModelFreeAuthoringDemonstration2022,wuAugmentedRealityInstruction2016,tainakaGuidelineToolDesigning2020,buttnerAugmentedRealityTraining2020,werrlichComparingHMDBasedPaperBased2018,houUsingAnimatedAugmented2013,blattgersteInSituInstructionsExceed2018,hendersonExploringBenefitsAugmented2011,alvesComparingSpatialMobile2019}), navigation (e.g.,~\cite{reitmayrCollaborativeAugmentedReality2004,zollmannFlyARAugmentedReality2014,mulloniHandheldAugmentedReality2011,hertelAugmentedRealityMaritime2021}), healthcare (e.g.,~\cite{changIntuitiveIntraoperativeUltrasound2005,heinrichComparisonAugmentedReality2020,samsetAugmentedRealitySurgical2008,reitingerSpatialMeasurementsMedical2005}), and civil engineering (e.g.,~\cite{schoenfelderAugmentedRealityIndustrial2008, schallHandheldAugmentedReality2009,schallVirtualRedliningCivil2008,sareikaUrbanSketcherMixed2007}). While some may not strictly be classified as ``information visualization'', we argue that they are examples of situated visualization, as they display information in situ using graphics. We therefore tap into this wealth of AR research by deriving common patterns of how visualizations can be designed with respect to physical environments and referents. To the best of our knowledge, no work has yet attempted to extract such ``situated visualization idioms'', nor considered the design variables and constraints in designing situated visualizations. This viewpoint is also shared in recent surveys~\cite{shinRealitySituationSurvey2023, bressaWhatSituationSituated2022}. 

In this work, we propose a system of design patterns for situated visualizations (see Figure~\ref{fig:teaser} for examples). We first search the AR literature, identifying a catalog of 10 common patterns of situated visualizations which cover a wide range of existing designs. We use these patterns as a foundation to devise six design dimensions which help describe and categorize situated visualization. We also identify five key constraints which influence the design of situated visualization. Lastly, we discuss several critical research directions that arose from our internal discussions throughout the course of this work.

To summarize, we  present insights from juxtaposing and fusing existing designs published in the AR and visualization research communities, which leads to three contributions:
\begin{enumerate}[noitemsep]
\item 10 design patterns which describe and summarize common techniques for situated visualization in the literature
\item An analysis of six design dimensions based on the patterns, revealing their underlying characteristics and organizational principles
\item A set of five guidelines organized by the real-world constraints that must be considered by situated visualization designers
%\item Key research directions necessary to attain a comprehensive understanding of the design of situated visualization
\end{enumerate}

% -----------------------------------------------------------------------------------------
\section{Background}
% -----------------------------------------------------------------------------------------

In this section, we provide a brief introduction to the emerging research fields of immersive and situated visualization. We summarize the previous work on design guidelines and design spaces for situated visualization, and highlight their limitations relevant to our work.

\subsection{Immersive Visualization and Analytics}

Immersive analytics has been described by Marriott et al.~\cite{marriottImmersiveAnalytics2018} as ``the use of engaging, embodied analysis tools to support data understanding and decision making.'' It is now increasingly common to use VR or AR head-mounted displays (HMD) for immersive analytics, especially as newer generations of hardware make the technology cheaper and more accessible \cite{fonnetSurveyImmersiveAnalytics2021}. 
VR is commonly used by researchers as a test bed to explore novel visualization and interaction techniques (e.g.,~\cite{cordeilImAxesImmersiveAxes2017,leeDeimosGrammarDynamic2023}), as they are not constrained by the comparatively lower performance and field of view (FOV) of present day AR glasses.

Nevertheless, a sizable amount of immersive analytics research has used AR displays---be they optical or video see-through---in their prototypes and user studies. Common reasons are to enable tangible or tactile interaction~\cite{satriadiTangibleGlobesData2022,smileyMADEAxisModularActuated2021}, to use the physical environment as scaffolding for arrangement and placement of virtual content~\cite{luoWhereShouldWe2022,leeDesignSpaceData2022}, or to allow collaborators to physically see each other~\cite{smileyMADEAxisModularActuated2021}. Yet, these works omit any semantic relationship between the data being visualized and the physical environments in which they are situated.

\subsection{Situated Visualization and Analytics}

White and Feiner~\cite{whiteSiteLensSituatedVisualization2009} defined \textit{situated visualization} as a ``visualization that is related to and displayed in its environment.'' In his dissertation, White~\cite{whiteInteractionPresentationTechniques2009} later described three key characteristics that a situated visualization must have: (1) The data in the visualization is related to the physical context; (2) the visualization is based on the relevance of the data to the physical context, and (3) the display and the presentation of the visualization lie in the physical context. The relationship to physical context (or, reality) is what sets situated visualization apart from regular (immersive) visualization.

Willett et al.~\cite{willettEmbeddedDataRepresentations2017} expanded the definition of situated visualization by formalizing the concept of \textit{physical data referents}, which are ``the real-world entities and spaces to which data corresponds.'' Whether or not a visualization is considered ``situated'' is determined by its spatial proximity to the referent to which it is related. They distinguished three types of visualizations: \textit{non-situated visualization} which is displayed in a different location than its referent (if any), \textit{situated visualization} which is located in the same location as its referent, and \textit{embedded visualization} which is both in the same location and spatially aligned with its referent such that they are viewed simultaneously. 

ElSayed et al.~\cite{elsayedSituatedAnalytics2015} proposed \textit{situated analytics}, which combines situated visualization with visual analytics~\cite{thomasIlluminatingPathResearch2005} to enable highly interactive, real time exploration and analysis of data that is related to physical referents. Note that not all situated visualizations need to be used in analytics contexts. As such, we do not consider the analytical processes nor interactions that situated visualizations can facilitate in this work. We refer to a very recent survey by Shin et al.~\cite{shinRealitySituationSurvey2023} on situated analytics for discussion around this topic.

The two aforementioned definitions of situated visualization by White and Feiner~\cite{whiteSiteLensSituatedVisualization2009} and Willett et al.~\cite{willettEmbeddedDataRepresentations2017} have been identified by Bressa et al.~\cite{bressaWhatSituationSituated2022} to be the most prevalent in the current literature. They also make several observations that are relevant to our work. First, the majority of situated visualization work uses AR. Our work also investigates situated visualization from an AR perspective, as it offers the widest design space possible while being largely unrestricted by physical constraints. Second, situatedness has commonly been associated with spatial relationships between visualizations, objects, and locations. We also consider situatedness from a predominantly spatial perspective to align with previous work. Third, there has been limited research on the design of the situated visualizations themselves. In this work, we describe design patterns to allow situated visualization designers to better understand and create situated visualizations.

\begin{figure*}[!ht]
\centering
 \includegraphics[width=\linewidth]{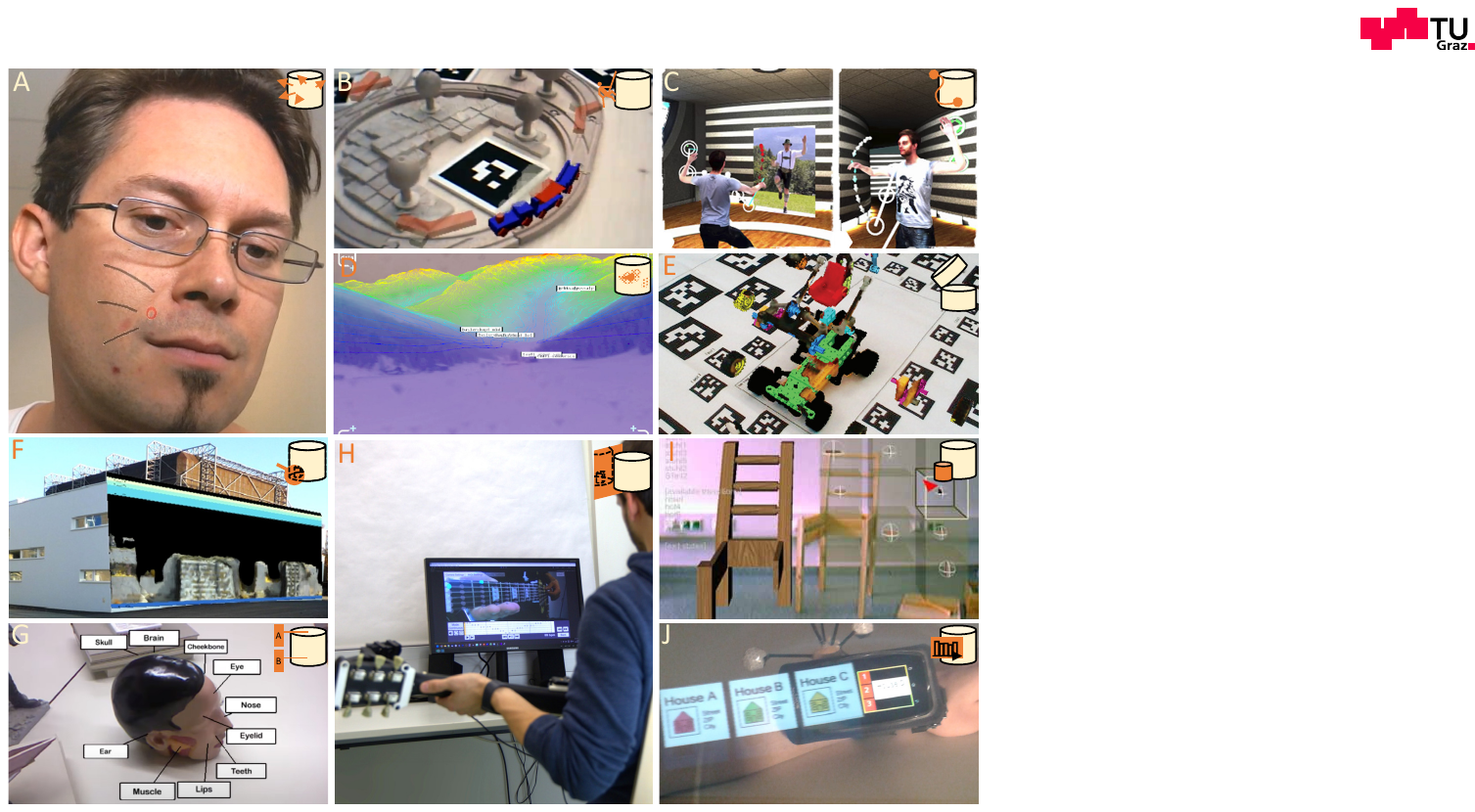}
\caption{Examples of our 10 situated visualization patterns, with image references: (A) glyph~\cite{mohrRetargetingVideoTutorials2017}; (B) ghost~\cite{wagnerMassivelyMultiuserAugmented2005}; (C) trajectory~\cite{yuPerspectiveMattersDesign2020}; (D) decal~\cite{veasExtendedOverviewTechniques2012}; (E) morph~\cite{kalkofenExplosionDiagramsAugmented2009}; (F) lens~\cite{zollmannInteractive4DOverview2012}; (G) label~\cite{tatzgernHedgehogLabelingView2014}; (H) mirror~\cite{skreinigARHeroGenerating2022}; (I) proxy~\cite{ledermannAPRILHighlevelFramework2005}; (J) panel~\cite{grubertMultiFiMultiFidelity2015}. For a description of these examples, see Section~\ref{sec:design-patterns}.}
\label{fig:allpatterns}
\vspace{-5mm}
\end{figure*}

\subsection{Design Spaces of Situated Visualization}

A limited number of works consider the design space of situated visualization. Willett et al.~\cite{willettEmbeddedDataRepresentations2017} characterize the two foundational classes (situated and embedded) and provide numerous design considerations for situated visualization, some of which we incorporate in this work. Most related to our work is that by Bach et al.~\cite{bachDrawingARCANVASDesigning2017} on AR-CANVAS. The ``CANVAS'' refers to the physical world into which AR elements are embedded. The term is also an acronym that describes several constraints imposed by the AR-CANVAS: Context-data, Artifact, Navigator, Visualization, Activity, and Scene. They then describe a design space that needs to be considered when using the AR-CANVAS: visual marks (what shapes are used), location (where the marks are placed), dimensionality (is the visualization 2D or 3D), orientation (how is the visualization oriented), visibility (when can the visualization be seen), styling (what does the visualization look like), and visualization design (what visualization idiom is used). While their taxonomy has comprehensive coverage of the problem space, the paper does not discuss the implications of the design space in much detail, nor does it provide practical examples of applying the design space.

A few other papers describe design spaces related to situated environments. Kawsar et al.~\cite{kawsarExploringDesignSpace2011} described four broad design cardinals for situated glyphs: \textit{what} information to present; \textit{how} is it presented; \textit{where} the glyphs are placed; and \textit{when} the glyphs are displayed. Of course, their work focused mainly on the design of individual glyphs, and did not consider other possible visualization types. Ens et al.~\cite{ensEtherealPlanesDesign2014} describe a design space for 2D information views in 3D mixed reality environments. They propose ``perspective'' (egocentric or exocentric) and ``movability'' (whether a display is fixed in the main frame of reference or not) as design variables, among others. These concepts are related to the user's frame of reference, which we also explore in our work. Danyluk et al.~\cite{danylukDesignSpaceExploration2021} describe a design space for \textit{worlds in miniature} (WIM)~\cite{stoakleyVirtualRealityWIM1995}. As a WIM replicates a large geographical space (or ``world'') in a miniature representation, they can technically be considered as proxies of physical referents~\cite{willettEmbeddedDataRepresentations2017, satriadiProxSituatedVisualizationExtended2023}. This observation is especially true when the WIM is used in-situ with the actual world which they represent. However, they mainly consider ways in which the WIM itself can be represented and interacted with (i.e.,~the presentation of the referent itself), rather than how the WIM can be annotated or supplemented with data representations. Morais et al.~\cite{moraisShowingDataPeople2022} include situatedness as part of their design space of anthropographics, with persons being the physical referents. They distinguish between four levels of situatedness from low to maximum, which can loosely correspond to the classes of situated visualization from Willett et al.~\cite{willettEmbeddedDataRepresentations2017}. Bowman et al.~\cite{bowmanInformationrichVirtualEnvironments2003} discuss information-rich virtual environments, which combine VR with information visualization by annotating the synthetic objects in the VR environment with visualizations generated from abstract information. Dillman et al.~\cite{dillmanVisualInteractionCue2018} study a related topic, namely how games use visual interaction cues (which can be seen as a form of visualization) and how these cues could be used in AR.

Despite not specifically relating to situated visualization, other research in immersive analytics has explored ways of spatially embedding visualizations with physical objects, even if they do not strictly fall under the definition of situated visualization. Satriadi et al.~\cite{satriadiTangibleGlobesData2022} devised a design space for how a tangible globe (i.e.,~the referent) can be used to explore geospatial data presented in AR either overlaid on top of or next to the globe. The work on augmented displays by Reipschl{\"a}ger et al.~\cite{reipschlagerPersonalAugmentedReality2021} demonstrated how 2D visualizations presented on wall displays can be augmented with AR visualizations in the spatial regions in front. Lagner et al.~\cite{langnerMARVISCombiningMobile2021} described techniques for extending data visualizations on tablets with AR views, such as providing alternative views next to the tablet or superimposing 3D visualizations directly on top of it.

In summary, previous work has not attempted a holistic overview of the visual design of situated visualization. They either have a narrow scope targeting a specific type of situated visualization~\cite{kawsarExploringDesignSpace2011, moraisShowingDataPeople2022}, examine situated visualization from a non-visual perspective~\cite{bressaWhatSituationSituated2022}, or do not sufficiently explain the creation and application of their design space~\cite{bachDesignPatternsData2018}. Our work seeks to fill this gap by providing an overview of the design of situated visualization based on a literature search combined with our experiences in AR designs.

% -----------------------------------------------------------------------------------------
\section{Design Pattern Catalog for Situated Visualization}
\label{sec:design-patterns}
% -----------------------------------------------------------------------------------------

This work was born from our realization that there exist a number of distinct \textit{situated visualization patterns} that are popular and recurring in the literature. Design patterns in general are commonly used in visualization (e.g.,~\cite{bachDesignPatternsData2018, javedExploringDesignSpace2012}) as they serve as standardized approaches to common tasks. For our purposes, situated visualization patterns are functionally similar to visualization idioms~\cite{munznerVisualizationAnalysisDesign2014}. That said, we avoid calling our patterns ``idioms'' to prevent confusion, especially as situated visualizations may employ conventional visualization idioms directly~\cite{bachDrawingARCANVASDesigning2017}. Therefore, our patterns serve as a high-level set of designs that we believe are representative of present-day situated visualization.

\subsection{Scope}
\label{sec:scope}

We approached the goal of formalizing a pattern catalog for situated visualization by examining how others have designed situated visualizations.
To broaden our scope, we include works that do not strictly fall under the definition of information visualization~\cite{munznerVisualizationAnalysisDesign2014, cardReadingsInformationVisualization1999}, such as visualizations of 3D data (e.g.,~\cite{kalkofenExplosionDiagramsAugmented2009, feinerCutawaysGhostingSatisfying1992}) or of instructional messages (e.g.,~\cite{tainakaGuidelineToolDesigning2020}). This extended scope lets us benefit from the ample research conducted outside of the visualization community.
We also include works that are not strictly \textit{situated}, so long as the referent is some physical object (e.g.,~in tangible and tactile interaction \cite{langnerMARVISCombiningMobile2021, satriadiTangibleGlobesData2022, reipschlagerDesignARImmersive3DModeling2019}) or is the environment itself (e.g.,~\cite{buschelMIRIAMixedReality2021}). This captures many adjacent works in AR-based immersive analytics that were not intended as situated.

To keep our scope concise, we only consider works that use AR. As such, we consider neither physicalizations~\cite{willettEmbeddedDataRepresentations2017}, nor the design or creation of the physical referents themselves. We also focus only on referents that are physically co-located with the user, since including (virtual) proxy referents would introduce myriad factors that require their own consideration~\cite{satriadiProxSituatedVisualizationExtended2023}. 
Moreover, we concentrate on the visual representation of situated visualization, and not on its interactivity. We acknowledge that interactivity is essential for interrogating the data and assisting in sensemaking~\cite{thomasIlluminatingPathResearch2005, yiDeeperUnderstandingRole2007,elsayedSituatedAnalytics2015}. Yet, since this work presents a first attempt to lay out the design space of situated visualization, we decided to constrain the scope to the visual representation only.

\subsection{Methodology}

We started with an existing corpus of over 2000 papers that were collected by the authors over more than two decades of research. From this corpus, 457 papers were selected by filtering the title, keywords and abstract (if available) with the terms ``augmented reality'', ``mixed reality'', and ``situated''.
The most senior author made a first pass over these papers, performing a free coding effort with the goal of identifying patterns. The two other authors performed a second pass over the papers, assigning them to the existing patterns. One author was assigned two-thirds of the corpus, with the other taking the remaining third. During this process, the senior author actively clarified any ambiguities and uncertainties with the other two authors.

Any papers which were deemed irrelevant were discarded. This included books, survey/position papers, and AR tracking/rendering papers. However, the latter were retained if they included a practical visualization intended for users. Conceptual designs of AR visualizations were included. Papers were assigned to multiple patterns where applicable, regardless of whether the patterns were used together simultaneously or used in different instantiations. At any time, it was permitted to add additional papers or refine the pattern catalog. All authors then discussed results and agreed on a final pattern catalog (Figure~\ref{fig:allpatterns}). The final corpus consists of 293 papers.

\subsection{Patterns}

We now describe a catalog of 10 main patterns based on our survey, many of which have specializations (marked in \textit{italics}). Note that these patterns are intended as categorizations used to describe common situated visualizations. Thus, certain instantiations may fall under the definition of multiple patterns, as can be seen in Figure~\ref{fig:teaser}. Please see our supplementary material for further breakdown and discussion on the use of each pattern in the literature.

\begin{wrapfigure}{R}{0.11\textwidth}
\centering
\vspace{-4mm}
\hspace{-4mm}
\includegraphics[height=1.8cm]{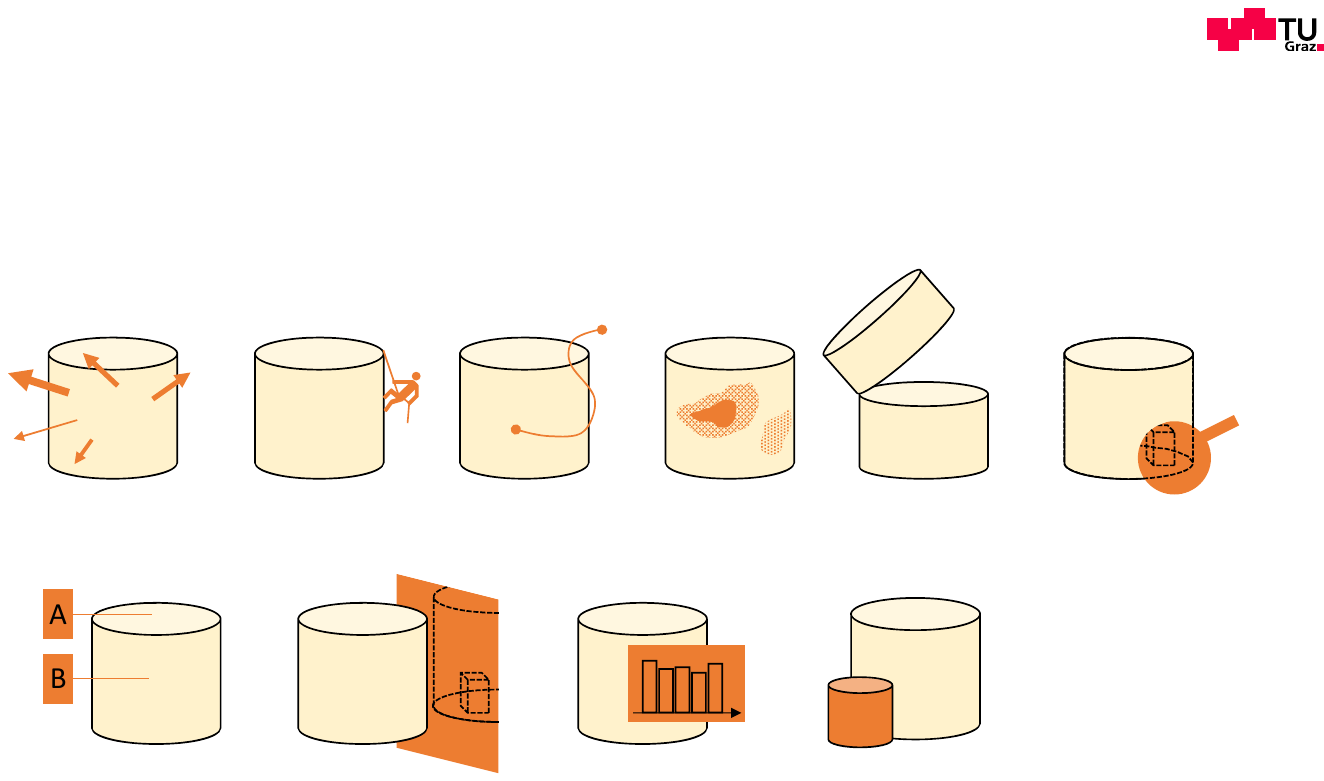}
\vspace{-5mm}
\end{wrapfigure}
\paragraph*{Glyph.} 
A glyph is simply a visual encoding of some information associated with a referent, which is placed so it is touching the referent. If a referent has several data elements associated to its parts or even to all points on its surface or interior, multiple glyphs can be used. In other words, glyphs can appear as singletons, in a sparse arrangement or even in a denser arrangement, provided that the resulting clutter is tolerable. As an example, consider sparse measurements (temperature, pressure) related to points on a surface. Glyphs can also draw attention to specific parts of the referent to assist in some task, such as in Figure~\ref{fig:allpatterns}A with make-up instructions \cite{mohrRetargetingVideoTutorials2017}. They can also provide additional visual cues to the user, such as to aid depth and orientation perception in 3D manipulation \cite{chintamaniImprovedTelemanipulatorNavigation2010}.

\begin{wrapfigure}{R}{0.11\textwidth}
\centering
\vspace{-4mm}
\hspace{-3mm}
\includegraphics[height=1.8cm]{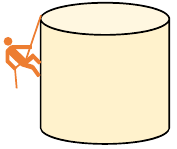}
\vspace{-3mm}
\end{wrapfigure}
\paragraph*{Ghost.} 
A ghost is a virtual object primarily intended to complement the scene, i.e.,~it resembles a (non-existing) physical referent. The difference between a ghost and a glyph is that the ghost's appearance is primarily that of a conventional object and not a visual encoding of particular data quantities. In other words, the ghost resembles a physical object that could be part of the scene. However, it must be noted that the boundary between glyphs and ghosts is not always precise, for example, if a free visual variable of the ghost object (e.g.,~its color) is used to encode some aspect of the data. Examples include planning layout in a factory~\cite{herrImmersiveModularFactory2018}, ghost cars on a race track showing the driving performance of past racers, or placement of virtual furniture in one's home with the latest IKEA app. Figure~\ref{fig:allpatterns}B shows ghost trains on real wooden tracks~\cite{wagnerMassivelyMultiuserAugmented2005}, and Figure~\ref{fig:teaser}, left, shows an outline indicating how to handle a tool.

\begin{wrapfigure}{R}{0.10\textwidth}
\centering
\vspace{-4mm}
\hspace{-3mm}
\includegraphics[height=2cm]{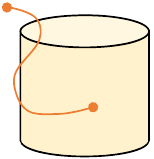}
\vspace{-3mm}
\end{wrapfigure}
\paragraph*{Trajectory.} 
A trajectory is a special kind of glyph that connects two or more endpoints. It can represent the motion of a virtual or real object, for example, in the form of a \textit{streamline}, or in a simpler case, as a directional \textit{arrow} (which is also a glyph). A trajectory can also be used as a \textit{visual link} (especially, a \textit{leader line}) which makes the logical connection between two entities, for example a referent and its label floating in free space. Another use case is an \textit{outline}, which highlights a referent without occluding it. The appearance of a trajectory will typically be in the form of a line or curve. As examples, consider human motion paths~\cite{buschelMIRIAMixedReality2021} or planned trajectories for drones~\cite{zollmannFlyARAugmentedReality2014}. Figure~\ref{fig:allpatterns}C shows trajectories depicting suggested dance motions~\cite{yuPerspectiveMattersDesign2020}.%, and Figure~\ref{fig:teaser}, middle, shows virtual markings at a construction site.

\begin{wrapfigure}{R}{0.09\textwidth}
\centering
\vspace{-5mm}
\hspace{-2mm}
\includegraphics[height=1.9cm]{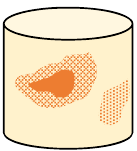}
\vspace{-5mm}
\end{wrapfigure}

\paragraph*{Decal.} 
A decal is a modification of the appearance of a physical surface. It is best explained as texture mapping applied to the surface of a real object. The texture contains a visual encoding of abstract variables. For example, color could be used to overlay a \textit{heatmap} or to simply \textit{highlight} the referent.
Decals have been used to show aggregated stay durations of physical movements in a room~\cite{luoPearlPhysicalEnvironment2023}, highlight which piano keys make a chord~\cite{barakonyiAugmentedRealityAgents2005}, construction progress of buildings~\cite{zollmannInteractive4DOverview2012}, and isolines on terrain~\cite{veasExtendedOverviewTechniques2012}. A decal serves a similar purpose as a set of glyphs placed on a surface, but it does so in a dense manner, assigning a data value to every covered point on the surface. Compared to glyphs, such dense coverage lets the observer associate values with surface points directly without having to mentally interpolate values observed on the nearest glyphs first. However, this directness comes at the price of increased occlusion; at the limit, the surface is no longer directly visible, since it is fully covered by the decal. Therefore, it is common to apply the decal only in selected areas or make it partially transparent to strategically reveal the referent underneath. Figure~\ref{fig:allpatterns} shows height isocontours and a heat map encoding water levels overlaid on a mountainscape~\cite{veasExtendedOverviewTechniques2012}.

\begin{wrapfigure}{R}{0.09\textwidth}
\centering
\vspace{-3mm}
\hspace{-3mm}
\includegraphics[height=2.1cm]{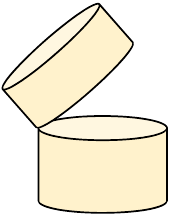}
\vspace{-5mm}
\end{wrapfigure}
\paragraph*{Morph.} 
A morph (colloquial for metamorphosis) is a modification of the physical objects in a scene (including, but not limited to referents). After reshaping, the pose, size or shape of objects appears differently. The most common purpose of applying a morph is to resolve clutter at the expense of unwanted occluders: For example, the occluders can be \textit{exploded}~\cite{kalkofenExplosionDiagramsAugmented2009}, \textit{shrunk}~\cite{sandorEgocentricSpacedistortingVisualizations2009} or \textit{diminished} (i.e.,~made invisible)~\cite{moriSurveyDiminishedReality2017}. \textit{Deformations} (e.g.,~bending a referent toward the observer) can instead be used to achieve quite the opposite, namely, to better present the referent to the observer~\cite{veasExtendedOverviewTechniques2012}. Figure~\ref{fig:allpatterns}E shows an explosion diagram applied to a physical toy car~\cite{kalkofenExplosionDiagramsAugmented2009}.

\begin{wrapfigure}{R}{0.11\textwidth}
\centering
\vspace{-5mm}
\hspace{-4mm}
\includegraphics[height=2cm]{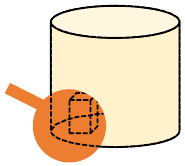}
\vspace{-5mm}
\end{wrapfigure}
\paragraph*{Magic Lens.} 
A magic lens (or lens, for short) is a visualization tool that changes the appearance of a portion of the user's FOV. Magic lenses were originally proposed for 2D interfaces~\cite{bierToolglassMagicLenses1993} and later for 3D interfaces~\cite{viega3DMagicLenses1996}. In 3D, a lens can be flat (e.g.,~a planar polygon) or volumetric (e.g.,~a cube); the latter enables delivering the lens effect to the observer independently of the viewing direction. Lenses in the form of \textit{portals}, i.e.,~wormholes that lead to other places, are widely known in popular culture~\cite{schmalstiegSewingWorldsTogether1999}. The concept of lenses is also popular in AR, where the visualization is embedded in the direct view of reality, but constrained to the extent of the lens~\cite{kalkofenInteractiveFocusContext2007}. Arguably, the most popular use are \textit{X-ray} lenses which reveal the interior of the referent~\cite{feinerCutawaysGhostingSatisfying1992}. The limited spatial extent of the lens can be used to compensate for the occlusion or clutter introduced by the embedded visualization. A common pattern uses a lens of finite extent which can be grabbed and manipulated by the observer~\cite{looser3DFlexibleTangible2007}, so clutter is avoided by spatiotemporal multiplexing: The lens covers different spatial areas at different times. Figure~\ref{fig:allpatterns}F shows a lens indicating internal construction data on the side of a physical building~\cite{zollmannInteractive4DOverview2012}.

\begin{wrapfigure}{R}{0.12\textwidth}
\centering
\vspace{-5mm}
\hspace{-2.5mm}
\includegraphics[height=2cm]{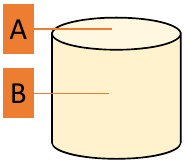}
\vspace{-6mm}
\end{wrapfigure}
\paragraph*{Label.} 
A label is a pattern typically used in an embedded view. Labels are intended to supply additional information to referents. Labels inform the observer about aspects of the physical environment that would not be conventionally accessible. Despite this rather straightforward use case, labels may well be the killer feature for the success of AR in the near future. The widespread availability of AR labels may have an impact on everyday life which could be as profound as the impact of spontaneous Wikipedia search on everyday conversations today. Physical labels (e.g.,~name tags) are usually placed directly on referents, because a physical support is required. Labels connected to referents by leader lines are commonly found in textbook illustrations, but less often in reality (although participants holding up billboards on sticks at political rallies do fit the definition). In AR, we are free to place labels anywhere in space, and such freedom is usually invested in avoiding occlusions. For example, labels indicating popular tourist destinations can be placed floating above the city's skyline~\cite{grassetImagedrivenViewManagement2012}. Labels placed directly on object surfaces should rather be categorized as decals or glyphs (see above). Unlike their real-world counterparts, AR labels have the unique advantage that they do not have to remain in a static relationship with their referent. The existence, placement and appearance (e.g.,~scale) of a label can be dynamically optimized for perception and for fitting the observer's information needs (cf.~the ``head'' model in Figure~\ref{fig:allpatterns}G). Label content is not restricted to text, although text may be the most common use case. An arrangement of textual labels in AR may be considered a form of text visualization. However, labels are by no means limited to textual content. Each label could contain an independent visualization (e.g.,~a bar chart) or other non-trivial visual content, such as a video loop. As an example, consider the anatomical labels on the head model \cite{tatzgernHedgehogLabelingView2014} shown in Figure~\ref{fig:allpatterns}G.

\begin{wrapfigure}{r}{0.11\textwidth}
\centering
\vspace{-5mm}
\hspace{-2.6mm}
\includegraphics[height=2.1cm]{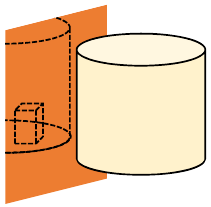}
\vspace{-8mm}
\end{wrapfigure}
\paragraph*{Virtual Mirror.}
A virtual mirror shows a reflected view of a part of the environment. The mirror plane separates the real world from the reflected one, with the latter showing a visual encoding applied to the reflected copy, rather than a faithful duplicate of the original. The most common configuration of a virtual mirror mimics a dressing mirror, set up to show a reflection of the person standing in front of it (e.g.,~for anatomy studies~\cite{borkBenefitsAugmentedReality2019}). Other configurations have been demonstrated as well, such as augmentations on selfie images (e.g.,~\cite{mohrRetargetingVideoTutorials2017,riglingYourFaceVisualizing2023}).
%handheld facial mirrors or mirrors at the back of a workbench showing a mirrored workpiece~\cite{yamaguchiVideoAnnotatedAugmentedReality2020}, as seen in Figure~\ref{fig:allpatterns}. 
Virtual mirrors are not necessarily restricted to front-facing cameras reflecting physical objects; in some cases, a virtual mirror reflects an embedded virtual view~\cite{bichlmeierVirtualMirrorNew2009} rather than a real one. Figure~\ref{fig:allpatterns}H shows a live-video mirror with glyph-like overlays for learning the guitar \cite{skreinigARHeroGenerating2022}. 

\begin{wrapfigure}{R}{0.10\textwidth}
\centering
\vspace{-5mm}
\hspace{-3mm}
\includegraphics[height=1.9cm]{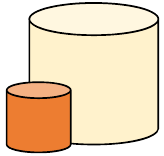}
\vspace{-5mm}
\end{wrapfigure}
\paragraph*{Proxy.} 
A \textit{proxy} is a pattern which involves a visualization near the user which resembles, in some or all aspects of its visual representation, a referent which is farther away. In terms of content and appearance, proxies may resemble individual objects or entire scenes, either in 3D (i.e.,~a \textit{WIM}) or reduced to a 2D \textit{map} representation (such as Google Maps). The proxy may also include a visual copy of the physical scene, for example, by using a hand-held AR display with video see-through or by using image-based rendering~\cite{witherIndirectAugmentedReality2011}. A \textit{portal} can be considered a proxy specialization as well. A proxy is frequently used to bridge larger distances or deal with out-of-reach objects. It also helps in dealing with scale discrepancies, letting the user handle huge or tiny objects that would be difficult to handle otherwise. Figure~\ref{fig:allpatterns}I uses a 1:1 proxy to indicate the next step in an assembly sequence~\cite{ledermannAPRILHighlevelFramework2005}.

\begin{wrapfigure}{R}{0.11\textwidth}
\centering
\vspace{-6mm}
\hspace{-3mm}
\includegraphics[height=1.8cm]{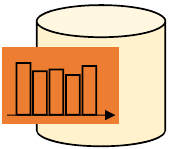}
\vspace{-6mm}
\end{wrapfigure}

\paragraph*{Panel.} 
A panel is a conventional visualization shown together with the real scene. The content of the panel is semantically linked to a referent, but lacks a geometric relationship to the physical environment. Oftentimes, a panel will use a 2D visual encoding, which is suitable for abstract data without any geometric aspects. However, a panel can also be a volumetric region in which a 3D visualization is presented (for example, a 3D graph or a 3D scatter plot). Since a panel does not have a spatial context in the scene, it can be placed arbitrarily in the scene, and we do not care in which coordinate frame the panel is placed. Panels can be set up as heads-up display, they can be attached to the user's body, or they can be placed in the environment. Figure~\ref{fig:allpatterns}J shows a view attached to the forearm~\cite{grubertMultiFiMultiFidelity2015}; and Figure~\ref{fig:teaser}, right, shows histograms overlaid on sections of a library. 
%A panel may also be migrated between coordinate frames without significant changes to its operation. For example, a panel may be head-referenced while a person is moving, but world-referenced when the same person is seated.

\subsection{Pattern Usage in the Corpus}
\label{sss:coding-results}
\begin{figure}
     \centering
     \includegraphics[width=\linewidth]{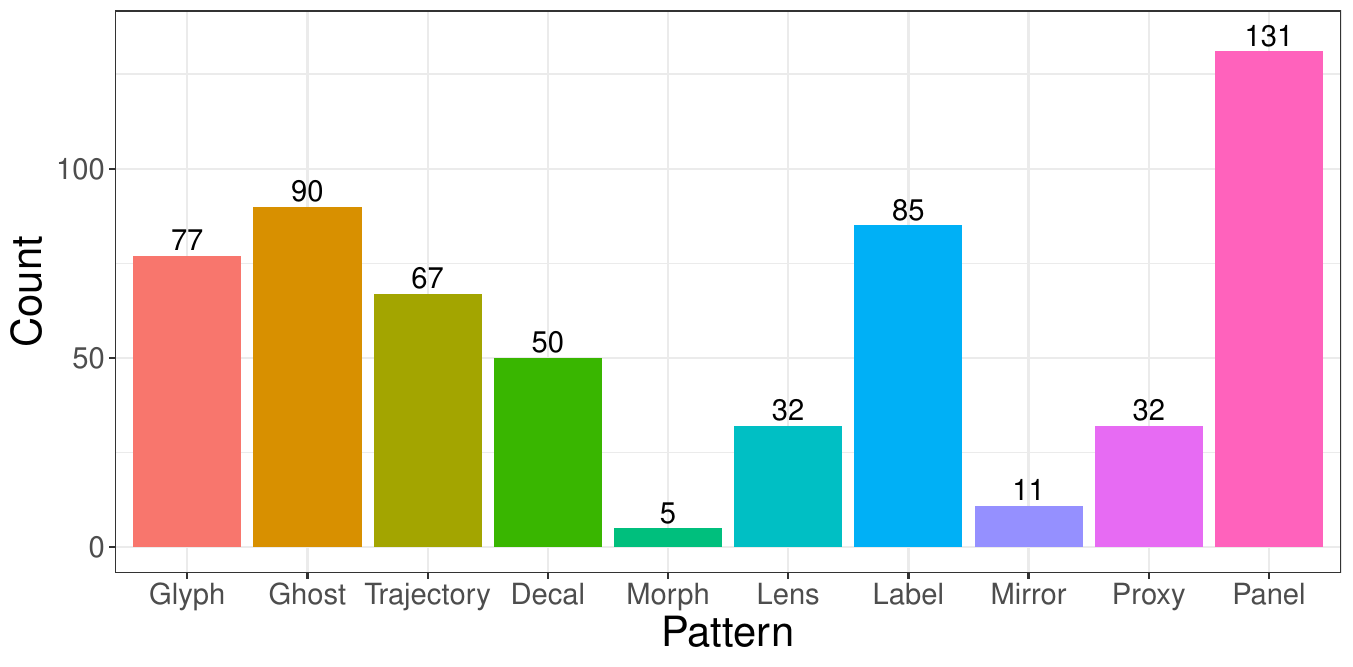}
     \vspace{-5mm}
    \caption{The number of occurrences of each pattern identified in our survey of 293 papers. Each paper may be assigned to more than one pattern.}
    \label{fig:survey-results}
    \vspace{-5mm}
\end{figure}

Figure~\ref{fig:survey-results} shows the frequency of each pattern in our corpus of 293 papers. Each paper was allowed to be assigned to more than one pattern. We briefly summarize insights into the use of these patterns.

\textbf{Panels} are the most common pattern due to their flexibility and familiarity. They can display any information the same way as a conventional display monitor can, ranging from text instructions (e.g.,~\cite{qianScalARAuthoringSemantically2022,wuAugmentedRealityInstruction2016}) to data visualizations (e.g.,~\cite{prouzeauCorsicanTwinAuthoring2020,fleckRagRugToolkitSituated2022}).
\textbf{Ghosts} are mainly used to supplement the physical scene, such as using avatars to add further context to spatiotemporal movement data (e.g.,~\cite{luoPearlPhysicalEnvironment2023,buschelMIRIAMixedReality2021,caoGhostARTimespaceEditor2019}). As AR can easily render 3D objects, it is no surprise that ghosts are popular.
\textbf{Glyphs} and \textbf{trajectories} are typically for visual guidance whilst performing a physical task, such as to indicate which parts of the referent to touch and manipulate (e.g.,~\cite{blattgersteInSituInstructionsExceed2018,skreinigARHeroGenerating2022}). They are also common for navigation (e.g.,~\cite{zollmannFlyARAugmentedReality2014,reitmayrCollaborativeAugmentedReality2004}), to indicate both waypoints as well as a route to follow.
\textbf{Labels} follow their standard usage of providing information at discrete parts of their referent (e.g.,~\cite{tatzgernHedgehogLabelingView2014,grassetImagedrivenViewManagement2012}). They can, however, also dynamically update their content based on user input (e.g.,~\cite{reitingerSpatialMeasurementsMedical2005}).
\textbf{Decals} generally spatially encode continuous field data overlaid on the referent (e.g.,~\cite{luoPearlPhysicalEnvironment2023,veasExtendedOverviewTechniques2012}). Decals can also provide further information/context as guidance for some task on the referent's surface, such as a virtual grid or target (e.g.,~\cite{alvesComparingSpatialMobile2019,heinrichComparisonAugmentedReality2020}). 
\textbf{Proxies}, while themselves comparatively less common than other patterns, are predominantly used in the form of WIMs and top-down maps. WIMs in particular can be used for not just navigation, but also for exploration of large or distant 3D objects (e.g.,~\cite{hoangAugmentedViewportAction2010}).
\textbf{Lenses}, as expected, are most frequently used to see the internal structure of referents as X-rays (e.g.,~\cite{kalkofenInteractiveFocusContext2007,feinerCutawaysGhostingSatisfying1992}). This does mean their applications are niche, hence their lower overall usage. However, there exist alternative uses of lenses, such as to see directly through objects as though they were transparent (e.g.,~\cite{lilijaAugmentedRealityViews2019,mendezImportanceMasksRevealing2009}).
\textbf{Mirror} and \textbf{morph} are the two least used patterns. For mirrors, this is likely due to the preference of embedding visualizations onto the referents directly; mirrors are mainly used for tasks that require awareness of the user's own body (e.g.,~\cite{riglingYourFaceVisualizing2023,andersonYouMoveEnhancingMovement2013}). For morphs, the act of altering the referent's appearance may make it difficult to keep awareness of its true physical structure, thus making it impractical to manipulate while morphed.

For a longer form description and analysis of the use of the patterns in the corpus, please refer to the supplementary material.

% -----------------------------------------------------------------------------------------
\section{Design Dimensions of Situated Visualization} 
% -----------------------------------------------------------------------------------------
\label{design-space}

\begin{figure*}[!ht]
    \centering
    \includegraphics[width=0.95\textwidth]{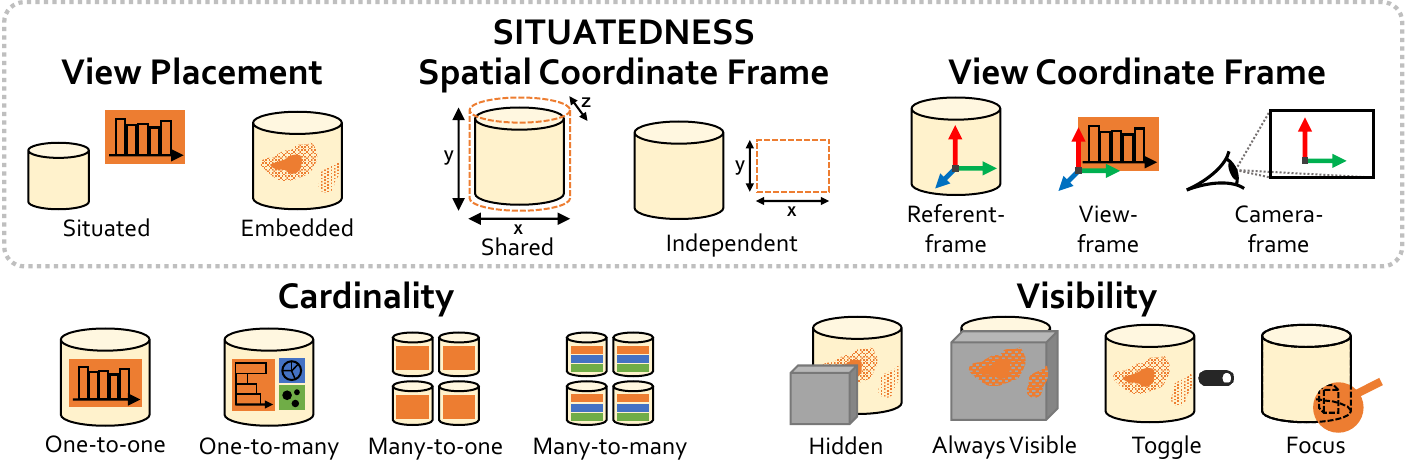}
    \caption{Overview of our design dimensions. \textit{View Placement:} Where is the view placed relative to its referent? \textit{Spatial Coordinate Frame:} How does the view encode data with its internal spatial structure? \textit{View Coordinate Frame:} How does the view move when the referent or camera moves? \textit{Cardinality:} What is the numerical relationship between referent(s) and view(s)? \textit{Visibility:} In what conditions can the view be seen?
    }
    \label{fig:design-space}
\vspace{-5mm}
\end{figure*}

We now propose several design dimensions of situated visualization.
Due to the sheer variety of visualizations in the literature, our patterns served as a starting point from which we derived an initial set of dimensions. We iterated on these dimensions throughout many discussions between the authors, and validated them using a random sampling of published works. We also leveraged our own prior experience working in both the AR and immersive analytics research fields.

The pattern catalog covers most, if not all, AR visualizations, but we do not claim that our resulting design space is in any way complete or comprehensive. Our scope is the same as that described in Section~\ref{sec:scope}. Consequently, there are three things to keep in mind. First, we do not focus specifically on information visualization, and thus our design dimensions are data- and visualization-agnostic. Second, we focus strictly on the design of the situated visualization and not on the physical referent. In other words, we assume that the representation of the physical referents are fixed and outside of the situated visualization designer's control. We instead view the pre-existing make-up and properties of the referents as design constraints, which we discuss later in Section~\ref{sec:constraints}. Third, we consider situated visualization from an AR perspective. That said, certain dimensions may theoretically apply to non-AR visualization as well (physicalization). We now describe our five design dimensions (Figure~\ref{fig:design-space}) in turn. 

\subsection{Situatedness: View Placement}
\label{sss:view-placement}

The term ``situatedness'' in the literature is strongly associated with the spatial relationship between referents and visualizations~\cite{bressaWhatSituationSituated2022}. This spatial relationship is of utmost importance, as it can influence users' ability to interpret and understand visualizations~\cite{polysRoleDepthGestalt2011}. We consider situatedness in terms of three design dimensions, which are intrinsically linked together due to their effect on spatial positioning. These are described in Sections~\ref{sss:view-placement}, \ref{sss:spatial-coordinate-system}, and \ref{sss:view-coordinate-frame}. Note that, in the case of multiple views, each may have a different form of situatedness.

Willett et al.~\cite{willettEmbeddedDataRepresentations2017} originally described two classes of situated views: situated and embedded. The two describe the level of spatial indirection between the view and the referent. While we re-use their terminology, we consider the two classes mainly from the perspective of \textit{view placement}. That is, the view is perceived to be located relative to its referent.

\paragraph*{Situated.} The view is in the same location as the referent, but is not perceived to be embedded, overlapping, or otherwise attached to the referent. Situated views can therefore usually stand on their own, and are comprehensible without needing to visually refer back to the referent. These patterns include mirror, proxy, and panel patterns, which are displayed externally from the referent.

\paragraph*{Embedded.} The view is perceived to be part of or attached to the referent. This impression may be caused by visual changes to the referent's geometry (morph) or surface (decal, lens). It may also be through visual augmentations that are aligned directly adjacent or close to the referent (glyph, ghost, trajectory, label).

\subsection{Situatedness: Spatial Coordinate Frame}
\label{sss:spatial-coordinate-system}

Views often encode data via their internal spatial structure. Such an encoding may be based on an abstract coordinate frame, such as mapping data to spatial positions as per information visualization. Alternatively, situated visualization may use a concrete coordinate frame, such as the position of navigational markers in the real world. In contrast to this internal coordinate frame, the referent itself presents its own (real-world) coordinate frame. We describe two options for how these \textit{spatial coordinate frames} (view and referent) may be used.
%\ds{talk about milgram congruence?}

\paragraph*{Shared.} The view follows the same coordinate frame as the referent. This mode can be thought of as the view using the spatial substrate of the referent, rather than a substrate defined by axes \cite{cardReadingsInformationVisualization1999}. Patterns include glyph, ghost, trajectory, and decal, which display information at discrete points or continuous regions of the referent (or even along its entire surface). Any information encoded in the view is semantically linked to the same position as on the referent itself. For example, glyphs can be embedded on a guitar's fretboard to indicate which exact notes were missed the most while practicing music scale exercises~\cite{heyenAugmentedRealityVisualization2022,skreinigARHeroGenerating2022}. When the view is instead situated, this would take the form of a proxy with a replicated geometric structure to its referent (e.g.,~\cite{satriadiTangibleGlobesData2022,stoakleyVirtualRealityWIM1995,danylukDesignSpaceExploration2021,tatzgernExploringDistantObjects2013,hoangAugmentedViewportAction2010}).

\paragraph*{Independent.} The view does not follow the same coordinate frame as the referent. This mode is similar to traditional information visualization, with it defining its own spatial substrate. The position or design of the view is therefore in some ways arbitrary, as it does not need to conform to the physical geometry of the referent. Simple examples are data visualizations that can be authored and freely positioned near their physical referents as panels (e.g.,~\cite{prouzeauCorsicanTwinAuthoring2020, fleckRagRugToolkitSituated2022}).

\subsection{Situatedness: View Coordinate Frame}
\label{sss:view-coordinate-frame}

The \textit{view coordinate frame} is the third and final design dimension related to situatedness. This dimension determines how the view is able to move when either the referent or the camera moves. The ``camera'' refers to the device that mediates the AR experience: typically a head-mounted or handheld display. View coordinate frame is not to be confused with the aforementioned spatial coordinate frame. The former is the \textit{external} coordinate frame of the entire view, whereas the latter is the \textit{internal} coordinate system used by the view's components. We describe three possible options for the view coordinate frame:

\paragraph*{Referent-frame.} The view follows its referent's coordinate frame. If the referent moves, the view moves along with it. In contrast, if the camera moves, the view appears to remain in place with its referent. Clearly, such views are suited for spatially tracked referents that are typically handheld (e.g.,~\cite{heyenAugmentedRealityVisualization2022,skreinigARHeroGenerating2022,satriadiTangibleGlobesData2022,rekimotoMatrixRealtimeObject1998}). Note that the view does not necessarily need to be embedded with the referent---the view can be merely situated and yet still be referent-fixed (e.g.,~as a \textit{side-by-side} view \cite{satriadiTangibleGlobesData2022}). Referent-fixed views (e.g.,~glyph, decal) can establish a stronger semantic connection with their referents due to having a common fate \cite{kohlerGestaltPsychologyIntroduction1947} as they both move together as one.

\paragraph*{View-frame.} The view establishes its own independent coordinate frame. As with referent-fixed, if the camera moves, the view appears to remain in place. But if the referent moves, the view does not follow either. In general, the view remains stationary in world-space (e.g.,~very large virtual mirrors \cite{borkBenefitsAugmentedReality2019,andersonYouMoveEnhancingMovement2013}). However, certain applications may allow the view-frame to be explicitly moved by the user. This mode allows users to further interrogate or organize the visualizations in their environment (such as in \cite{luoWhereShouldWe2022,leeSharedSurfacesSpaces2021}).

\begin{table*}[]
    \centering
    \renewcommand{\arraystretch}{1.65}
    \begin{tabular}{|c|p{3.6cm}|p{3.6cm}|p{3.6cm}|p{3.6cm}|} \hline
    \textbf{Frame}     & \textbf{Embedded + Shared} & \textbf{Embedded + Independent} & \textbf{Situated + Shared} & \textbf{Situated + Independent} \\ \hline
    \textbf{Referent}  
    & % Embedded, shared, referent
    \textit{Glyphs}, \textit{decals}, \textit{trajectories}, \textit{ghosts}, \textit{labels}, and \textit{morphs} which move with their referent \cite{hubenschmidSTREAMExploringCombination2021,langnerMARVISCombiningMobile2021,satriadiTangibleGlobesData2022,heyenAugmentedRealityVisualization2022,skreinigARHeroGenerating2022,rekimotoMatrixRealtimeObject1998}.
    & % Embedded, independent, referent
    Embedded \textit{panel} or \textit{proxy} which moves with its referent but uses independent axes \cite{smileyMADEAxisModularActuated2021,elsayedSituatedAnalytics2015,tongExploringInteractionsPrinted2022}.
    & % Situated, shared, referent
    Situated \textit{proxy} which follows the movement of its referent~\cite{satriadiTangibleGlobesData2022}.
    & % Situated, independent, referent
    Situated \textit{panel} which follows the movement of its referent \cite{ensUpliftTangibleImmersive2021}.
    \\ \hline
    \textbf{View}      
    & % Embedded, shared, view
    Standalone \textit{lens} or \textit{mirror} that can move independently from the referent \cite{kalkofenInteractiveFocusContext2007,looser3DFlexibleTangible2007}.
    & % Embedded, independent, view
    Standalone \textit{ghosts} of unit visualizations contextualized by familiar environments \cite{assorExploringAugmentedReality2023}.
    & % Situated, shared, view
    \textit{Proxy} of the referent that can move independently from the referent \cite{stoakleyVirtualRealityWIM1995,tatzgernExploringDistantObjects2013,hoangAugmentedViewportAction2010,cruzAugmentedRealityApplication2019}.
    & % Situated, independent, view
    Standalone \textit{panel} displaying information about the referent \cite{fleckRagRugToolkitSituated2022,prouzeauCorsicanTwinAuthoring2020,whitlockHydrogenARInteractiveDataDriven2020}.
    \\ \hline
    \textbf{Camera}    
    & % Embedded, shared, camera
    \textit{Labels} which automatically position around the referent in the HUD \cite{urataniStudyDepthVisualization2005,tatzgernHedgehogLabelingView2014,grassetImagedrivenViewManagement2012}.
    & % Embedded, independent, camera
    \textit{Panel} that updates its content based on which referent is in the camera's FOV \cite{fleckRagRugToolkitSituated2022}.
    & % Situated, shared, camera
    \textit{Proxy} of the referent in the HUD in the form of a top-down map \cite{dillmanVisualInteractionCue2018,veasExtendedOverviewTechniques2012,cruzAugmentedRealityApplication2019}.
    & % Situated, independent, camera
    \textit{Panel} presenting information on the user's HUD \cite{reitmayrCollaborativeAugmentedReality2004, tainakaGuidelineToolDesigning2020,hendersonAugmentedRealityPsychomotor2011,zhuAuthorableContextawareAugmented2013,mulloniHandheldAugmentedReality2011}.
    \\ \hline
    \end{tabular}
    \caption{Twelve possible configurations of the three situatedness dimensions (view placement, spatial coordinate frame, view coordinate frame). For each configuration, a possible visualization is given with corresponding pattern(s) and a non-exhaustive set of example references.}
    \label{tab:sit_table}
\vspace{-3mm}
\end{table*}

\paragraph*{Camera-frame.} The view follows the coordinate frame of the camera's FOV (i.e.,~screen space), commonly referred to as the heads-up display (HUD). Such views are typically configured to always face the camera to ensure that they are constantly readable. Whether or not the view moves based on the angle between the camera and the referent is up to the designer. A common approach is to automatically position the view so that it appears to be on top of or next to a referent (e.g.,~using labels \cite{urataniStudyDepthVisualization2005,tatzgernHedgehogLabelingView2014,grassetImagedrivenViewManagement2012}). If stationary, the view remains in place regardless of movement (e.g.,~panels on a HUD \cite{reitmayrCollaborativeAugmentedReality2004,hendersonAugmentedRealityPsychomotor2011,zhuAuthorableContextawareAugmented2013,mulloniHandheldAugmentedReality2011}). Alternatively, it may be beneficial to alter the visible content based on movement. For instance, the view (or parts of it) may only become visible when viewed within a specific region of the user's FOV (i.e.,~a lens \cite{looser3DFlexibleTangible2007,hoangAugmentedViewportAction2010,kalkofenInteractiveFocusContext2007}).

The three design dimensions---view placement, spatial coordinate frame, view coordinate frame---form 12 possible configurations of situatedness. Table~\ref{tab:sit_table} provides a concrete example of each combination.

\subsection{View Cardinality}

Physical environments can easily consist of more than a single referent. Likewise, a situated visualization can be made up of more than a single view. The numerical relationship between the two---known as the view cardinality---is a variable that can be controlled by the designer. We describe four possible cardinal relationships between the number of referents and the number of views. Note that the chosen cardinality need not be uniform for all referents and views. Different referents and views may follow different cardinalities, and some referents may not have any views associated with them at all.
 
\paragraph*{One-to-one.} The referent has a single view associated with it. This mode is the most basic and recognizable form of situated visualization and can be used by any pattern.

\paragraph*{One-to-many.} The referent has multiple views associated with it. This mode may occur in the form of multiple (coordinated) views~\cite{robertsStateArtCoordinated2007} which encode multi-dimensional data about the referent. For example, multiple panels with different visualizations can display information about a grocery store product \cite{elsayedSituatedAnalytics2015}. Alternatively, multiple patterns can be combined together. As examples, consider a panel showing temperature and pressure data while the user mimics the actions of refueling a car that is represented as a ghost \cite{whitlockHydrogenARInteractiveDataDriven2020}, or decals overlaid on the terrain showing spatio-temporal data with labels marking discrete points of interest \cite{veasMobileAugmentedReality2013}.

\paragraph*{Many-to-one.} Multiple referents are used together to comprise a single view. The simplest case has a single view that aggregates and displays information about multiple referents at once. These views are generally situated panels that use information visualization idioms. A more unique case is given when graphical marks are spread across multiple referents, with the totality of these marks forming the overall view. As such, this cardinality is useful in providing an overview of the data across all referents. For example, Guarese et al.~\cite{guareseAugmentedSituatedVisualization2020} place labels on each chair in a classroom which encode a specific data dimension using color (e.g.,~airflow, visibility). These labels collectively provide a spatial distribution of the room's environmental conditions.

\paragraph*{Many-to-many.} Multiple referents are used to represent multiple views. This mode can also be seen as an extension of many-to-one. Multiple values can be encoded on each referent to provide an overview of multiple data collections and of multidimensional information of each individual referent themselves. While not an AR example, Willett et al.~\cite{willettEmbeddedDataRepresentations2017} demonstrate this in an embedded data physicalization of conference attendees and their badges. All badges can be viewed from afar to get a composite overview of the makeup of the group via color encodings, while each badge can be viewed up close to see the attendee's name, role, affiliation, etc.

\subsection{Visibility}

Situated visualizations are, by definition, placed in the real world. There are likely situations where a view occludes an object of interest---be it the referent itself or another part of the environment. Therefore, views may need to be made hidden (or visible) under specific conditions. We describe four possible design options, some of which are based on the \textit{visibility} design decision proposed by Bach et al.~\cite{bachDrawingARCANVASDesigning2017}.

\paragraph*{Hidden.} The view becomes hidden when it is occluded and remains visible otherwise. This mode is arguably the most straightforward approach, as it mimics the behavior of real-world objects. Alternatively, the view may become hidden when the referent itself is occluded, be it any part or all of it.

\paragraph*{Always Visible.} The view is always visible, regardless of whether the referent itself is visible to the user. Such a setting is vital for the X-ray lens pattern, which is common in building management, where cables and pipes are commonly hidden and out of view \cite{schallHandheldAugmentedReality2009,schallVirtualRedliningCivil2008}. 

\paragraph*{Toggle.} The visibility of the view is manually toggled on and off by the user. This mode provides the user with fine-grained control over when and how the view is seen and used. Of course, its actual usability depends on the interaction technique used to toggle. Instructional panels in situated applications commonly use toggles to show/hide information, especially when following a sequence of steps (e.g.,~\cite{tainakaGuidelineToolDesigning2020,hendersonAugmentedRealityPsychomotor2011,blattgersteInSituInstructionsExceed2018,mohrRetargetingVideoTutorials2017}). The morph pattern is another example where a toggle can be beneficial, such as for temporarily revealing more information about the referent's structure in an explosion diagram \cite{kalkofenExplosionDiagramsAugmented2009}.

\paragraph*{Focus.} The view is made visible when the referent is somehow considered to be in ``focus''. Bringing referents and their views in focus can be an explicit action performed by the user, such as positioning a magic lens \cite{looser3DFlexibleTangible2007} or aiming a flashlight \cite{ridelRevealingFlashlightInteractive2014,ferdousWhatHappeningThat2019}. This action may also be implicit, such as when the user picks up, touches, or approaches the referent (e.g.,~\cite{ensUpliftTangibleImmersive2021,hendersonExploringBenefitsAugmented2011}). As we do not consider interaction in this work, however, further design considerations are not discussed here.

\section{Constraints and Guidelines} 
% -----------------------------------------------------------------------------------------
\label{sec:constraints}

We have now extracted design dimensions from the patterns, which let us classify the patterns and argue about the qualities of concrete designs derived from these patterns. However, unlike a desktop visualization designed to fill a blank screen, a situated visualization may need to adhere to the constraints imposed by its physical environment. These constraints can represent a significant obstacle in making the situated visualization effective in achieving its communicative and informative goals. Since the constraints are imposed by the real world, the situated visualization designer may have little to no influence over them, and thus needs to work within the given situation.

The constraints can be broadly categorized into technological constraints and physical constraints. Technological constraints are those borne from the hardware and software used to facilitate the situated visualization. For our purposes, these constraints primarily derive from the capabilities of the AR display and the sensors that collect the data to be used in the visualization. In contrast, physical constraints are those borne from the referents and the environment in question. In the following, we discuss the most important constraints from both categories, and describe guidelines on how to handle recurrent design challenges emerging from the constrained situation.

\subsection{Extent of World Knowledge}
\label{sss:ewk}

The seminal paper by Milgram and Kishino~\cite{milgramTaxonomyMixedReality1994} proposed the \textit{virtuality continuum}, which organizes mixed reality displays based on the relative amounts of real and virtual stimuli they contain. As a second, lesser known contribution, the paper also introduces three separate sub-continua: extent of world knowledge, reproduction fidelity, and extent of presence metaphor. Of the three, the \textit{extent of world knowledge} (EWK) is the one most relevant to situated visualization. It describes the amount of real-world information used in the AR display. On one end, nothing is known about the real world. On the other end, everything is known about the real-world, including its objects, their locations, their status, and so on.

Generally speaking, obtaining more complete knowledge of the environment (i.e.,~increasing the EWK) is beneficial, since it widens the possibilities of situated visualization and the scenarios which they can be used. Of course, acquiring such complete knowledge of the world may be too costly or impractical, especially out in the wild. Therefore, we can safely assume that the EWK is primarily perceived as a constraint by the visualization designer. As a simple example, a visualization cannot be spatially embedded with its referent if the system cannot recognize the presence of the referent in the first place.

From the end-user's perspective, EWK determines how ``smart'' a situated visualization appears. 
%As society becomes more reliant on technology, these smart visualization systems with high EWK may very well play an integral role in our everyday lives and decision-making. 
Conversely, a designer may intentionally limit the EWK utilized, even if it is otherwise available. Consider a florist arranging a bouquet with the assistance of a situated analytics application~\cite{elsayedSituatedAnalytics2015}. The application may know about the availability of flowers in stock, their storage location, product quality, client preferences, and so on. The application may then be able to calculate optimal bouquets and instruct the florist on how to arrange them outright. Alternatively, the application might instead let the florist decide how to arrange the bouquets based on their real-world perception and human judgment, merely providing data views to aid in the florist's decision-making. Thus, EWK acts as a one-way constraint: it only limits the scope of situated visualization if EWK is low (the user must interact or search more), and does not if it is already high.

\textbf{Guidelines}: If possible, maximize EWK to strongly ground the visualizations in the real world. If EWK is low, one can substitute a lack of real-world context by more user interaction (e.g.,~search for keywords instead of pointing to referents). To avoid too much tedious low-level interaction, it may be necessary to rely on a ``browsing'' interface mode with minimal interaction.

\subsection{Location Awareness}

Location awareness is the dimension along which we classify the frequency with which location (place and space) awareness of referents is acquired or updated.  Among the aspects that belong to the ``world knowledge'', location awareness plays a crucial role. Without it, an AR application would have to guess about the presence of referents, making it impossible to build sophisticated situated or embedded visualizations. The location awareness considered here only concerns the referents, not the update rate of the camera transform applied to the user's view of the scene (i.e.,~the self-tracking of the user). Even the problem of solely tracking the referents can become quite challenging if there are many of them, or if the referents are rapidly moving in and out of the current location.
We distinguish between \textit{static}, \textit{discrete} and \textit{continuous} location awareness. Similarly to EWK, location awareness is a one-way constraint, with \textit{static} having the most limitations and \textit{continuous} having the least. 

\paragraph*{Static Location Awareness.}
Applications with \textit{static} location awareness often acquire location information offline, storing it in an immutable database. Situated visualizations read this information and cross-reference it with their own location tracking. For example, a building information modeling (BIM) system may store the location of every item in the inventory of a large building (windows, doors, central heating, power sockets, network plugs, and so on). The identity and location of these inventorized items rarely changes. A situated visualization would therefore only have to query the BIM database once to determine the location of each referent. Static is the most constraining form of location awareness (excluding the case of \textit{no} location awareness), as a visualization must trust that the static database is accurate, and the application logic cannot react to spontaneous movement of referents. As such, static awareness may only be viable in physical environments with fixed referent locations.

\paragraph*{Discrete Location Awareness.}
Applications with \textit{discrete} location awareness obtain the location of a referent on demand, one measurement at a time. This level of location awareness is usually enabled by a technical device, utilizing either manual scanning or an automatic but narrowly scoped discovery via computer vision (e.g.,~QR codes, fiducial markers). Such a feature allows the system to know when the user is in the same location as the referent (by virtue of scanning a target in a known location), and even the position of the referent in 3D space. Such an approach is useful if a designer wants situated visualizations to adapt to dynamic scenarios, but cannot afford high-fidelity spatial tracking of the referents. For example, the popular commercial Vuforia tracking library supports up to 100 simultaneous tracking targets, but a target must be seen up close to be recognized. Hence, it is not possible to observe even a fraction of the targets at once, and applications with more than a handful of targets are rare.

\paragraph*{Continuous Location Awareness.}
Applications with \textit{continuous} location awareness are the most demanding of the three. The locations of referents are tracked in real time, and visualizations are instantly updated to reflect these changes. Optical tracking systems can provide spatial awareness in six degrees of freedom and therefore facilitate embedded visualizations on referents. For example, Uplift~\cite{ensUpliftTangibleImmersive2021} uses several situated visualizations on referents fitted with Vicon reflective infrared markers. These objects are continuously tracked by multiple stationary cameras, so the application can expect continuous location updates even if no user (with a head-worn camera) is looking at the referents. Such a condition relies on an external tracking infrastructure, while still being limited to a small number of referents. Even more disappointing is that, if multiple users stand close together, the tracking system may fail to deliver continuous updates due to line-of-sight occlusions. If the tracking accuracy of fast-moving referents is inadequate, it becomes hard to use embedded views with shared spatial coordinate frames, as they require good location awareness to accurately align the view and referent together. Instead, leaving the view as just situated (i.e.,~by using a proxy \cite{satriadiProxSituatedVisualizationExtended2023}) or using an independent coordinate system may be a better choice, since only the approximate location of the referent can suffice. 

\textbf{Guidelines}: Location awareness predominantly affects the placement of views---the lower the location awareness, the more situated and less embedded views should be used. Embedded views with shared spatial coordinate frames, due to their high precision requirements, are impractical when referents are moving and their spatial position cannot be continuously tracked. Possible remedies may be to affix the referents in predetermined locations or to design the interaction such that only one referent is moved at a time, thus opening the option of continuous tracking from a comparably inexpensive single-camera setup. 

%Even with state-of-the-art tracking, providing continuous awareness of many movable referents is very hard. For the foreseeable future, a situated visualization design should prefer solutions that only require static or discrete location awareness, or limit continuous location awareness to a single referent at a time.

\subsection{Referent Size} 

Referents that are too large or small in \textit{size} may be impractical for situated visualization---particularly for embedded views. Large referents may prove difficult to see the entire embedded view at once, while small referents may be obscured by the view. Small referents are also difficult for AR systems to track, thus exacerbating any tracking errors.

\textbf{Guidelines}: Situated views are generally unaffected by size constraints, so long as the user is still aware of the presence of the referent. Views with independent coordinate frames can thus be designed as per normal, typically using panels. Views with shared coordinate frames are most often proxies of the referent \cite{willettEmbeddedDataRepresentations2017}. If an embedded view is necessary, consider using a proxsituated configuration \cite{satriadiProxSituatedVisualizationExtended2023}, combining a proxy with other patterns in a one-to-many cardinality.

%An external, situated view can mitigate against large referents, while small referents can be handled by assigning views only to the whole referent and not to sections of it, such as points or areas on the referent. An alternate solution, especially for referents that are even further toward the extremes (e.g.,~microscopic), is to use the proxy of the referent \cite{willettEmbeddedDataRepresentations2017}. Unfortunately, proxies tend to diminish the physicality of AR. For an in-depth discussion of the subject, we refer to Satriadi et al.~\cite{satriadiProxSituatedVisualizationExtended2023}.

\subsection{Referent Density}
\label{sss:referent-density}

The \textit{density} of referents, including their number and arrangement, must also be considered. A \textit{low density} of referents is trivial to manage. As density increases, however, the space available for each individual visualization in the user's FOV decreases. Hence, if a \textit{high density} of referents must be accommodated, the designer needs to ensure readability of both the visualizations and the referents themselves.

\textbf{Guidelines}: Clutter from overlapping referents and views may be reduced by adjusting the view position, cardinality, and visibility. Situated views are the obvious solution, as they can be placed away from the referents, thus preventing occlusion from occurring. That said, embedded views that overlap graphics with their referent (e.g.,~decal, lens) may still be viable as they only occlude themselves, not other objects. Using different cardinalities can also help by simply reducing the number of views in the environment. Alternatively, many-to-one views may become more attractive, as the high density already provides an accessible overview to see all of the data. These can also aggregate the data of all referents into a single situated view, which is inherently highly scalable. Lastly, selective adjustment of the view visibility can reduce the amount of visual clutter shown at once, at the risk of hiding potentially important information items that are outside of the user's focus or are otherwise (accidentally) hidden.

\subsection{Navigational Requirements}
\label{sss:navigational-requirements}

\begin{figure}
    \centering
    \includegraphics[width=\linewidth]{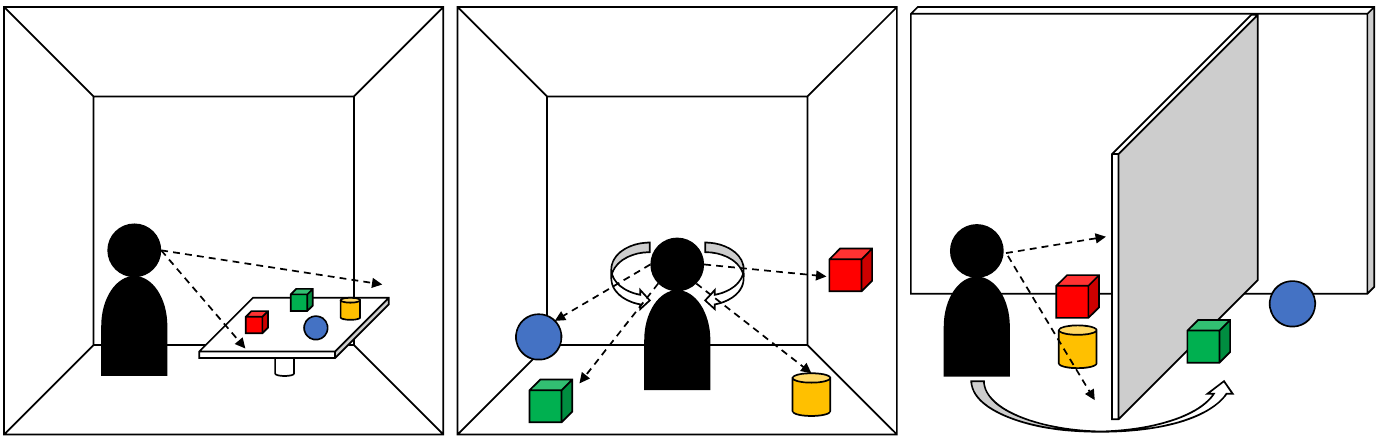}
    \vspace{-4mm}
    \caption{The layout of referents and environment affects the navigational requirements.
    Left: A narrow field of regard, the user can see all referents at once;
    middle: A wide field of regard, the user can see all referents from a single vantage point by rotating;
    right: Multiple fields of regard, the user can see all referents only by moving.
}
\vspace{-5mm}
\label{fig:envsize}
\end{figure}

When multiple referents are involved, we must also consider navigational requirements. This consideration is influenced by the relationship between the field of regard (FOR)---the region of space that contains the referents to be seen---and the field of view (FOV)---the region of space that can be seen at once (Figure~\ref{fig:envsize}). In the simplest case, the entire FOR can be covered by a single FOV (i.e.,~\textit{narrow field of regard}). This configuration is commonly associated with exocentric visualizations where the user looks inward toward a referent and its views. A slightly more complex case occurs when the FOR is larger than the FOV, but can still be covered through rotation (i.e.,~\textit{wide field of regard}). This configuration is associated with egocentric visualizations where the user looks outwards, as though they are surrounded by referents and views. The most complex case is when rotation alone is insufficient, and the user is required to physically move (i.e.,~\textit{multiple fields of regard}). This configuration is common in navigational applications or in environments where objects tend to be occluded due to their sheer scale (e.g.,~\cite{reitmayrCollaborativeAugmentedReality2004,cruzAugmentedRealityApplication2019}). Of course, the extent of the FOV is dependent on the AR device being used. A hand-held AR display necessitates greater movement to cover the same FOR than a HMD, for instance.

\textbf{Guidelines}: The effect on the design is dependent on the given task. If the task can be accomplished simply by looking at a single referent (or a group of nearby referents), the design consideration becomes similar to that of referent density (Section~\ref{sss:referent-density}). If the task requires context switching between multiple referents in a wide (or multiple) FOR, then some form of visual guidance may be required (e.g.,~\cite{joAroundplotFocusContext2011,langeHiveFiveImmersionPreserving2020}). If data of all referents must be seen simultaneously, it may be necessary to aggregate everything into a singular situated independent view, thus forgoing embedded views. If the referents are physically occluded, a lens may also be employed to provide X-ray vision in a specific focus area (e.g.,~\cite{lilijaAugmentedRealityViews2019,mendezImportanceMasksRevealing2009}), or the views may simply be set to be always visible.

% -----------------------------------------------------------------------------------------
\section{Discussion, Limitations, and Future Work}
\label{sec:discussion}
% -----------------------------------------------------------------------------------------

In this section, we discuss several promising directions for future work which arose during our survey and internal discussions. We believe that these are synergistic with the core contributions of this work. We also briefly detail the limitations of our work.% We then detail the limitations of our work.

\subsection{Design of Physical Referents}

A core assumption of this work is that the choice of physical referents is outside of the situated visualization designer's control. This assumption may not apply to some applications, such as museums, theaters, or edifices being newly constructed. By choosing the size, mobility, density or even the type of referents used, the designer may be able to optimize both the referent and visualization to best showcase their data. Note that this is markedly different from physicalization, as AR may still be used in conjunction with existing physical referents. 

\subsection{Interacting with Situated Visualizations}

While our design patterns and survey do not consider interactivity, this is an obvious direction to investigate further, given its focus in immersive analytics \cite{marriottImmersiveAnalytics2018,fonnetSurveyImmersiveAnalytics2021}. We can already identify and briefly describe several forms of interaction. For tasks involving a sequential series of steps, the system may automatically navigate forward when it detects that the user has performed the task (e.g.,~\cite{mohrTrackCapEnablingSmartphones2019,hendersonAugmentedRealityPsychomotor2011,stanescuModelFreeAuthoringDemonstration2022}), or require manual stepping by the user (e.g.,~\cite{lochComparingVideoAugmented2016,hendersonExploringBenefitsAugmented2011,yamaguchiVideoAnnotatedAugmentedReality2020}). For more open-ended tasks, the user may simply be able to pick up the referent and view it from different angles, which may also adjust the view and the data shown (e.g.,~\cite{ensUpliftTangibleImmersive2021,satriadiTangibleGlobesData2022}). Views may be toggled and manipulated directly (e.g.,~\cite{zhengSTAREAugmentedReality2022,prouzeauCorsicanTwinAuthoring2020}) or implicitly through motion of the user's own body (e.g.,~\cite{qianScalARAuthoringSemantically2022,ferdousWhatHappeningThat2019}). Of course, a more thorough analysis will be required to understand the possibilities of interaction.

\subsection{Tasks in Situated Visualization}

The task which the situated visualization is intended to aid will likely have a significant impact on its design. As mentioned above, whether the task is performed in a strict sequence or not influences the visibility rules of the visualization. The use of situated versus embedded views is also likely to be task-dependent, as the user may need to see the physical state of the referent in an obstructed manner. In Section~\ref{sss:navigational-requirements}, we allude to hypothetical tasks which require the user to see all referents simultaneously, thus affecting the cardinality and visibility rules. In short, there is clearly a strong influence on the design based on the given task. Such task taxonomies, common in the visualization literature (e.g.,~\cite{shneidermanEyesHaveIt1996,amarLowLevelComponentsAnalytic2005}), would fit naturally to situated visualization.

% Dieter: I propose to remove the limitations section, since it is too defensive ("hedge words", as ben said). Better say in conclusion that these things warrant further work

\subsection{Limitations}
Lastly, we want to acknowledge that this paper is not a comprehensive, systematic literature survey---nor was it intended to be. While we examine previous work to form our patterns and draw high-level observations (Section~\ref{sss:coding-results} and supplementary material), it is arguably insufficient to determine underexplored areas of the literature. For that, we refer to recent surveys conducted by others \cite{bressaWhatSituationSituated2022, satriadiProxSituatedVisualizationExtended2023,shinRealitySituationSurvey2023}. However, we believe that our contribution to knowledge---namely that of our design patterns, dimensions, and constraints---are capable of standing on their own, especially with the theoretical foundation of our survey.

%Lastly, we want to acknowledge that our work has several limitations. First, this paper is not a comprehensive, systematic literature survey---nor was it intended to be. Its primary purpose was to identify common design patterns in the AR and visualization literature, with it serving as a foundation for our design dimensions and constraints. The paper was \textit{not} initially intended to identify underexplored areas of the situated visualization literature. For that purpose, we refer to surveys that have been conducted recently by others \cite{bressaWhatSituationSituated2022, satriadiProxSituatedVisualizationExtended2023}. Moreover, the AR research field is vast, and even though our corpus contains \bl{x} papers there are clearly many more that we did not include.
%Second, we only focused on AR visualization and not physicalization, which forms a significant portion of the situated visualization landscape \cite{willettEmbeddedDataRepresentations2017,bressaWhatSituationSituated2022}. We believe that this warrants its own design space exploration, as a number of our design dimensions are clearly incompatible with physicalization (i.e.,~view coordinate frame, visibility, display form factor).

% -----------------------------------------------------------------------------------------
\section{Conclusion}
% -----------------------------------------------------------------------------------------

One of the grand challenges of immersive analytics~\cite{ensGrandChallengesImmersive2021} calls for designing guidelines for spatially situated visualizations. Our work aims to be one of many steps toward addressing this challenge. We did so by surveying the literature to understand how researchers have designed situated visualization. We summarize this in a catalog of 10 design patterns. From these patterns, we have extracted a set of six design dimensions for the categorization of situated visualizations. Moreover, we describe five of the most important real-world constraints that affect situated visualization design and propose patterns and ways of dealing with these constraints. We hope that our work provides not only design guidelines, but also a shared vocabulary for how researchers understand, describe, and investigate situated visualization.
%As future work, we hope to extend our work in the research directions discussed in Section~\ref{sec:discussion}. 
%, which otherwise serve to build on the limitations of this work. 
%Most importantly, we plan on investigating how tasks affect the design of situated visualization.

% if specified like this the section will be ommitted in review mode
\acknowledgments{%
This work was funded by the German Research Foundation (DFG) project 495135767 and the Austria Science Fund (FWF) project I 5912-N (joint Weave project), and partially supported by Germany's Excellence Strategy – EXC 2120/1 – 390831618 (DFG).}

% \newpage
%\bibliographystyle{abbrv-doi-hyperref}
\bibliographystyle{abbrv-doi-hyperref-narrow}
%\bibliographystyle{abbrv-doi}
%\bibliographystyle{abbrv-doi-narrow}

%\bibliography{references, references-old}
\bibliography{references-abbrv}

%% ^^^^^   FOR IEEE VIS, EVERYTHING HERE MAY BE INCLUDED IN THE    ^^^^^ %%
%% 2-PAGE ALLOTMENT FOR REFERENCES, FIGURE CREDITS, AND ACKNOWLEDGEMENTS %%

\appendix % You can use the `hideappendix` class option to skip everything after \appendix

\end{document}

%%%%%%%%%%%%%%%%%%%%%%%%%%%%%%%%%%%%%%%%%%%%%%%%%%%%%%%%%%%%%%%%
%%%%%%%%%%%%%%%%%%%%%% START OF THE PAPER %%%%%%%%%%%%%%%%%%%%%%
%%%%%%%%%%%%%%%%%%%%%%%%%%%%%%%%%%%%%%%%%%%%%%%%%%%%%%%%%%%%%%%%

\firstsection{Introduction}

\maketitle

This supplementary document provides further insights from our survey, which applied a codebook of 10 situated visualization design patterns to a corpus of 293 papers. Please refer to Sections 3.1 and 3.2 in the main manuscript for full details about the survey's scope and methodology.

It is important to reiterate that the survey primarily serves as the theoretical and literary basis for our 10 situated visualization design patterns. Thus, this supplementary document goes into further detail as to how each design pattern has been used in the literature, rather than identifying research gaps and opportunities. For more in-depth and comprehensive literature surveys, particularly for the identification of research gaps and opportunities, we refer to a recent 2022 survey by Bressa et al.~\cite{bressaWhatSituationSituated2022} on situated visualization, and a very recent 2023 survey by Shin et al.~\cite{shinRealitySituationSurvey2023} on situated analytics. Note that almost all in-text citations in this document are non-exhaustive.

\section{Corpus characteristics} \label{sec:corpus-characteristics}
The vast majority of papers in the corpus originate from the visualization and augmented reality (AR) research communities, including ISMAR, VR, TVCG, and CHI. It includes full papers as well as short papers, workshop papers, and posters, such as those from ISMAR-ADJUNCT and CHI EA.

\begin{figure}[h]
\centering
 \includegraphics[width=0.9\linewidth]{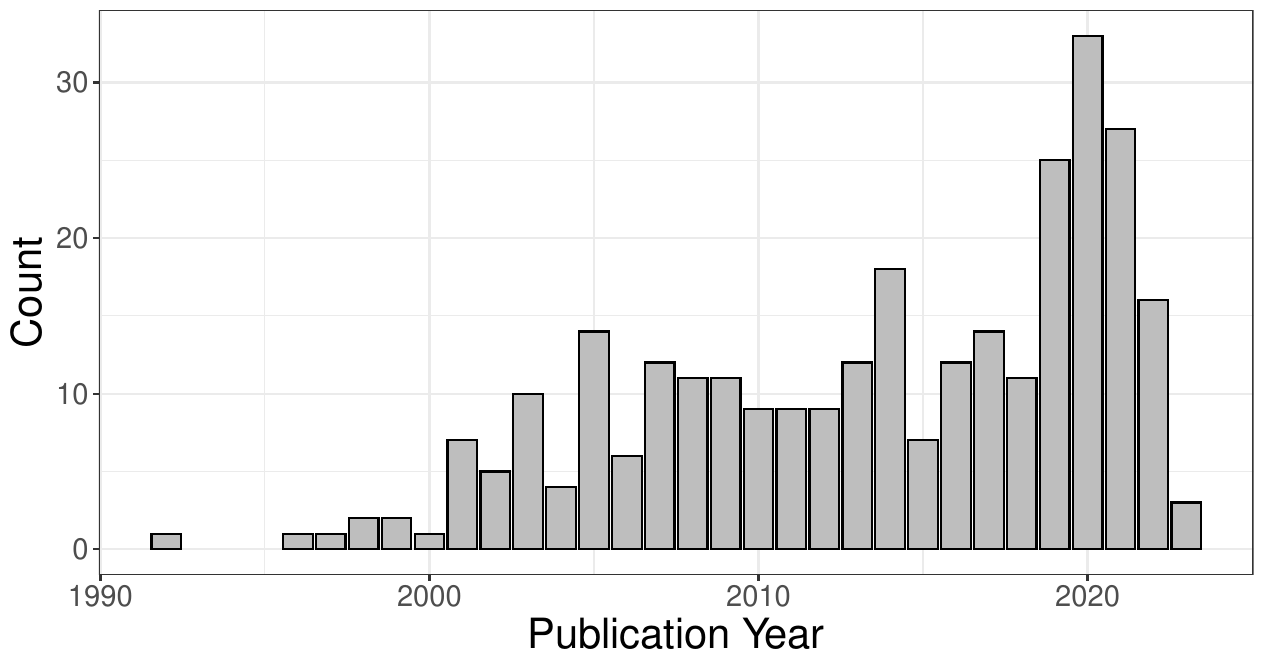}
\vspace{-2mm}
\caption{Number of papers published per year in our corpus.}
\label{fig:survey-yearly-histogram}
\vspace{-4mm}
\end{figure}

Figure~\ref{fig:survey-yearly-histogram} shows the number of papers that were published in each year in our corpus. As to be expected, there is a noted increase in papers roughly around 2019, in part due to the inclusion of numerous immersive analytics papers. It is important to stress however that these counts should not be used as an indication of research interest in situated visualization, due to our convenience sampling of papers as described in the original manuscript. Nevertheless, the corpus contains a fair balance of papers throughout the past 2+ decades, which we believe provides a consistent view of how the AR and visualization fields have used situated visualization. The only notable outlier in our corpus in terms of publication year is the paper on cutaways and ghosting by Feiner and Seligmann~\cite{feinerCutawaysGhostingSatisfying1992}, which was included due how well it demonstrates the so-called ``X-ray lens'' design.

\section{Popularity and usage of patterns}
\begin{figure}
\centering
 \includegraphics[width=\linewidth]{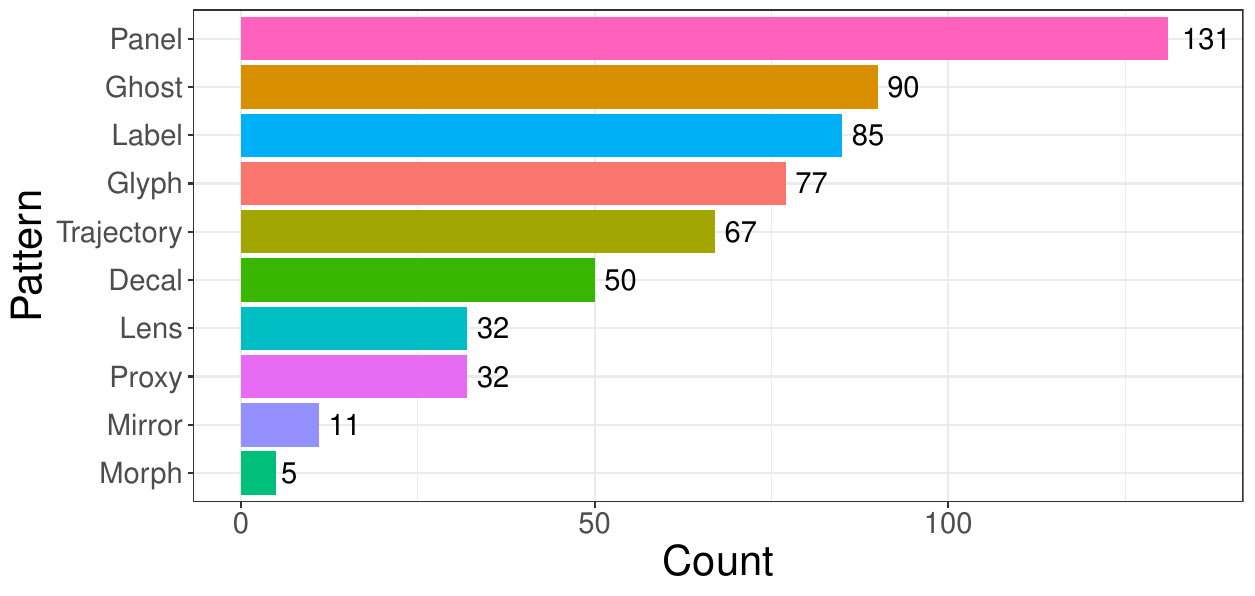}
 \vspace{-6mm}
\caption{Ranked frequency of each pattern in our survey. A paper may be assigned to more than one pattern.}
\label{fig:survey-pattern-counts}
 \vspace{-4mm}
\end{figure}

Table~\ref{tab:survey-table} lists the 10 patterns and provides references to the papers which were assigned to each of them. Figure~\ref{fig:survey-pattern-counts} provides a visual representation of the same data as the table, but sorted by frequency. In contrast to Section 3 of the original manuscript which sought to define the patterns, this section aims to provide further insight into the ways in which each pattern is used in our corpus (roughly sorted by frequency).

\textbf{Panels} are clearly the most used pattern in our corpus. This is not surprising, as it is a flexible, intuitive, and straightforward method of displaying information to the user. Panels are usually text-based (e.g., to show instructions/documents to the user \cite{kongTutorialLensAuthoringInteractive2021,qianScalARAuthoringSemantically2022,wuAugmentedRealityInstruction2016}, to show WIMP-style user interfaces \cite{leeEnhancingFirstPersonView2020,mohrMixedRealityLight2020,hubenschmidReLiveBridgingInSitu2022}). With the growth of immersive analytics, it is now common to use panels to show abstract data visualizations (e.g., \cite{prouzeauCorsicanTwinAuthoring2020,fleckRagRugToolkitSituated2022,kimVisARBringingInteractivity2017,tongExploringInteractionsPrinted2022}), as they are an obvious means to do so---particularly for 2D visualizations.

\textbf{Labels} are also a popular pattern as they are a simple and effective method of embedding information onto physical referents, with the majority of research using it as such. Labels can not only show static information (e.g., names of locations \cite{bellViewManagementVirtual2001,grassetImagedrivenViewManagement2012}), but also show dynamically updating values based on the state of the referent and/or user's input (e.g., for measurement \cite{reitingerSpatialMeasurementsMedical2005}, for 3D manipulation~\cite{kasaharaExTouchSpatiallyawareEmbodied2013}). As mentioned in the original manuscript, the placement of labels is a research challenge in and of itself, with myriad considerations such as clutter, label crossing, and overlap with other labels and salient areas \cite{grassetImagedrivenViewManagement2012}.

\textbf{Ghosts} are a unique pattern in that they are virtual objects that appear to be part of the actual physical scene, and thus serve to complement it. Given the interlink between AR and 3D graphics, it is unsurprising that ghosts are frequently used. Ghosts can stand on their own without being fixed to any physical referent (e.g., as an autonomously moving virtual agent~\cite{barakonyiAgentsThatTalk2004,barakonyiMonkeyBridgeAutonomousAgents2005}, for factory planning~\cite{herrImmersiveModularFactory2018}, to visualize full-body movements at discrete steps~\cite{caoGhostARTimespaceEditor2019}). Likewise, ghosts can also be affixed to a physical referent. A straightforward example are ghosts which indicate where on the referent a physical component should be attached in manual assembly (e.g., \cite{stanescuModelFreeAuthoringDemonstration2022,houUsingAnimatedAugmented2013,zogopoulosAuthoringToolAutomatic2022}). Another example is spatially superimposing the ghost onto the referent, such that the user believes they are instead holding the ghost (e.g., \cite{hettiarachchiAnnexingRealityEnabling2016,unluPAIRPhoneAugmented2021}). From an information visualization perspective, ghosts have been used to facilitate unit visualizations, with each data point being represented as a 3D object (e.g., \cite{chenMARVisTAuthoringGlyphBased2020,assorExploringAugmentedReality2023}). Ghosts can also provide further context to aid in the posthoc analysis of spatio-temporal data (e.g., an avatar representation of the user at a given point in time~\cite{luoPearlPhysicalEnvironment2023}, virtual representations of pre-existing objects~\cite{buschelMIRIAMixedReality2021}).

\textbf{Glyphs} are another common pattern. While glyphs can be used to encode some data directly on the physical referent (e.g., for immersive analytics \cite{satriadiTangibleGlobesData2022,buschelAugmentedRealityGraph2019,chenMARVisTAuthoringGlyphBased2020}), they are most popularly used to either highlight some part of the referent (e.g., using a floating marker \cite{blattgersteInSituInstructionsExceed2018,kasaharaJackInIntegratingFirstperson2014,leeEnhancingFirstPersonView2020}) and/or to indicate where some action should be performed (e.g., where to push \cite{kongTutorialLensAuthoringInteractive2021,skreinigARHeroGenerating2022}, where to drill/insert \cite{dennlerAugmentedRealityNavigated2021,kayaDynamicContentGeneration2021,houUsingAnimatedAugmented2013,heinrichComparisonAugmentedReality2020}). Glyphs can also be combined with the trajectory pattern as an arrow to denote a particular direction or path the action needs to be taken (e.g., rotate an object clockwise or counter-clockwise \cite{wuAugmentedRealityInstruction2016,tainakaGuidelineToolDesigning2020,caoExploratoryStudyAugmented2020}, look or move in the specified direction \cite{pustkaAutomaticConfigurationPervasive2011,mulloniUserExperiencesAugmented2011}). Note that these glyphs and trajectories may also be freely drawn by a user (e.g., during remote assistance as annotations \cite{nuernbergerInterpreting2DGesture2016,mohrMixedRealityLight2020,datcuHandheldAugmentedReality2016,gauglitzTouchRemoteWorld2014}, for sketching \cite{saquibInteractiveBodyDrivenGraphics2019,gasquesPintARSketchingSpatial2019,leivaRapidoPrototypingInteractive2021}).

\textbf{Trajectories}, as mentioned earlier, are typically used with glyphs to show directions, whether they be provided by the system or drawn by a user. They may indicate a path which an object (or the user) should directly follow (e.g., for drones~\cite{zollmannFlyARAugmentedReality2014}, for navigation~\cite{liShvilCollaborativeAugmented2014,reitmayrCollaborativeAugmentedReality2004,shinCAMARFutureDirection2009}). Trajectories are also obvious candidates for posthoc analysis of spatio-temporal data, particularly that of physical movements (e.g., \cite{buschelMIRIAMixedReality2021,hubenschmidReLiveBridgingInSitu2022,luoPearlPhysicalEnvironment2023}). Trajectories can also simply be used as regular trajectories and visual links as per information visualizations (e.g.,~\cite{reipschlagerPersonalAugmentedReality2021,satriadiTangibleGlobesData2022,langnerMARVISCombiningMobile2021}).
They may also be used to connect and draw links between multiple referents (e.g., in networks \cite{buschelAugmentedRealityGraph2019,ensIvyExploringSpatially2017,seigerHoloFlowsModellingProcesses2021,herbertUsabilityConsiderationsHand2020}, along different points of space \cite{reitmayrCollaborativeAugmentedReality2004,newmanWideAreaTracking2006,zhengLocationBasedAugmentedReality2019}), or between a referent and another visualization pattern (e.g., panels~\cite{zhuMechARspaceAuthoringSystem2022,whitlockHydrogenARInteractiveDataDriven2020}, labels~\cite{grassetImagedrivenViewManagement2012,tatzgernAdaptiveInformationDensity2016,rekimotoMatrixRealtimeObject1998}).

\textbf{Decals} are comparatively less common than other patterns. This may be for a number of reasons, from the need to accurately detect and track the referent's surface, to the possibility of the decal occluding the underlying referent. Nevertheless, decals have been used in a wide range of physical scales: large scale (e.g., in agriculture~\cite{kingARVinoOutdoorAugmented2005,zhengLocationBasedAugmentedReality2019}, building acceptance~\cite{zollmannInteractive4DOverview2012,schoenfelderAugmentedRealityIndustrial2008,sareikaUrbanSketcherMixed2007}, environmental monitoring~\cite{veasMobileAugmentedReality2013,veasExtendedOverviewTechniques2012}); medium scale (e.g., in rooms~\cite{luoPearlPhysicalEnvironment2023}); and small scale (e.g., on tabletops~\cite{alvesComparingSpatialMobile2019}, on human faces~\cite{riglingYourFaceVisualizing2023}).
Decals can show continuous field data (e.g., of physical movements~\cite{luoPearlPhysicalEnvironment2023}, of crop yields~\cite{kingARVinoOutdoorAugmented2005}), can be used similarly to panels embedded on the referent (e.g., affinity diagramming with post-it notes~\cite{subramonyamAffinityLensDataAssisted2019}), and can provide more visual guidance to facilitate a given task (e.g., a virtual grid on a table~\cite{alvesComparingSpatialMobile2019,chastineStudiesEffectivenessVirtual2008}, a virtual target~\cite{heinrichComparisonAugmentedReality2020}, overlaid textures to indicate selection~\cite{kasaharaExTouchSpatiallyawareEmbodied2013}).

\textbf{Proxies} are similar to ghosts in that they appear as virtual 3D objects, but dissimilar in that they are replications of the referent itself (i.e., digital twin). As such, proxies are used comparatively little as the user can oftentimes already see the referent itself. As stated in the original manuscript, proxies are generally used to visualize referents that are too large in scale or are difficult to acquire (e.g., buildings \cite{tatzgernDynamicCompactVisualizations2013,gruberCitySightsDesign2010}, human organs~\cite{kalkofenIntegratedMedicalWorkflow2006,schmalstiegStudierstubeAugmentedReality2002}). Because proxies are inherently virtual, they can be freely manipulated by the user without physical constraints for myriad purposes (e.g., to explore intricate parts of the referent~\cite{tatzgernExploringRealWorld2015,tatzgernExploringDistantObjects2013,guvenVisualizingNavigatingComplex2006}, for zooming~\cite{hoangAugmentedViewportAction2010}, for scaling~\cite{ledermannAbstractionImplementationStrategies2007}, for shopping and try-on \cite{joIoTARPervasive2019,cruzAugmentedRealityApplication2019}). A proxy can also be used to indicate the desired state that the referent should be in (e.g., after product assembly~\cite{lochComparingVideoAugmented2016,yamaguchiVideoAnnotatedAugmentedReality2020,houUsingAnimatedAugmented2013}). Of course, proxies are perhaps most commonly used for navigation (e.g., as 3D worlds-in-miniature~\cite{ledermannAPRILHighlevelFramework2005,prouzeauCorsicanTwinAuthoring2020,wagnerFirstStepsHandheld2003,mulloniHandheldAugmentedReality2011,reitmayrLocationBasedApplications2003,kalkuschStructuredVisualMarkers2002,newmanWideAreaTracking2006}, as 2D top-down maps~\cite{zhengLocationBasedAugmentedReality2019,speiginerEvolutionArgonWeb2015}).

\textbf{Magic Lenses} fulfill a niche in that they are mainly used to reveal the internal structure of referents. As with proxies (and to some extent morphs and mirrors), this typically requires a digital twin of the referent in order to accurately visualize and position the contents revealed by the lens. Lenses can be used when it is important to look inside of a referent (e.g., to understand structures without occlusion~\cite{kalkofenInteractiveFocusContext2007,mendezInteractiveContextdrivenVisualization2006,feinerCutawaysGhostingSatisfying1992}, for surgery~\cite{gsaxnerMarkerlessImagetoFaceRegistration2019,dennlerAugmentedRealityNavigated2021}, for maintenance~\cite{schallUrban3DModels2007,mohrRetargetingTechnicalDocumentation2015}). It can also be used to look directly through referents as though they were transparent (e.g., while driving~\cite{rameauRealTimeAugmentedReality2016}, for occluded interaction~\cite{lilijaAugmentedRealityViews2019,mendezImportanceMasksRevealing2009,eratDroneAugmentedHumanVision2018}).
Lenses may either reveal a portion of the referent when viewed through a specified region of space (e.g., \cite{kalkofenInteractiveFocusContext2007,looser3DFlexibleTangible2007,mendezInteractiveContextdrivenVisualization2006,krugCleARSightExploring2022}), or be always visible to the user (e.g., \cite{schallHandheldAugmentedReality2009,junghannsEmployingLocationawareHandheld2008}). A compelling form of lens is to utilize a flashlight metaphor for shining AR visuals onto referents (e.g., on cultural artifacts in museums~\cite{ridelRevealingFlashlightInteractive2014}, on human bodies~\cite{changIntuitiveIntraoperativeUltrasound2005}). Another compelling use of lens is to leverage a physical handheld lens to enable tangible interaction of AR objects \cite{krugCleARSightExploring2022}.

\textbf{Mirrors} are one of the two underutilized patterns. A likely reason is that mirrors require the user to look at a 2D plane that is separate from the original referent, increasing the level of spatial indirection which makes it harder to perform some physical task. That said, mirrors are perfectly suited in applications where the user needs to see their own face or body (e.g., \cite{mohrRetargetingVideoTutorials2017,riglingYourFaceVisualizing2023,borkBenefitsAugmentedReality2019,andersonYouMoveEnhancingMovement2013,yuPerspectiveMattersDesign2020}), just like an actual mirror. Mirrors can also be used in place of head-mounted AR displays (e.g., using projectors~\cite{benkoMirageTableFreehandInteraction2012,bimberAugmentedRealityBackProjection2000}, using a monitor~\cite{yamaguchiVideoAnnotatedAugmentedReality2020,skreinigARHeroGenerating2022,reitmayrWearable3DAugmented2001}), which may be beneficial if wearing a head-mounted display is impractical.

\textbf{Morphs} are the rarest pattern found in the literature. While manipulating the appearance of a referent can resolve issues like occlusion (e.g., explosion diagrams~\cite{kalkofenExplosionDiagramsAugmented2009}, diminished reality~\cite{moriSurveyDiminishedReality2017}), it may be problematic when the user still requires awareness of the physical structure of the referent---particularly when they need to touch and manipulate it. Therefore, most works that modify the appearance of a referent would keep to salient augmentations like ghosts, or preserve the outer structure of the referent while revealing the internal structure using a magic lens.

\section{Pattern usage over time}
\begin{figure}[th]
\centering
 \vspace{-3mm}
 \includegraphics[width=\linewidth]{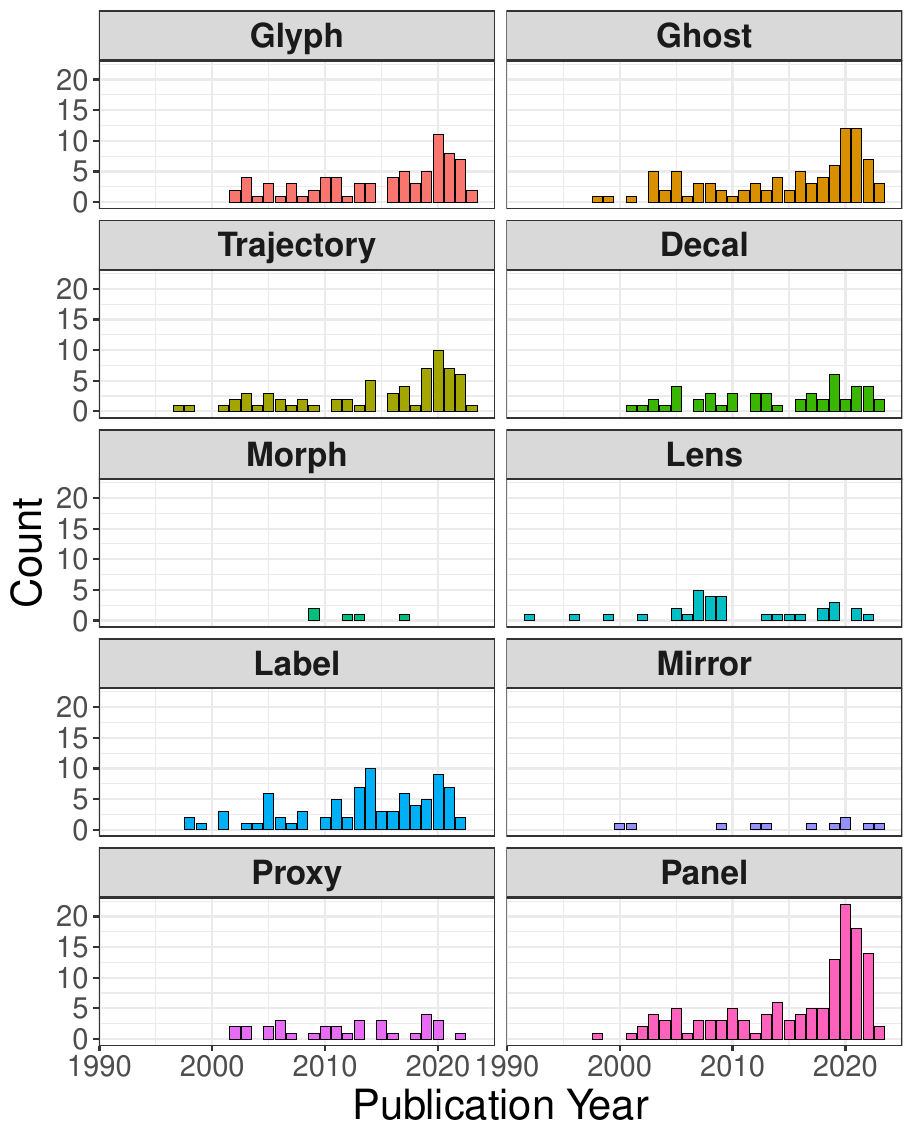}
 \vspace{-6mm}
\caption{Number of papers per year in our corpus, faceted by pattern.}
 \vspace{-3mm}
\label{fig:survey-yearly-faceted}
\end{figure}

Figure~\ref{fig:survey-yearly-faceted} shows the frequency that each individual pattern appeared in our corpus per year. Glyph, ghost, trajectory, label, and panel all clearly see a spike in usage in 2019 and beyond. Of course, this is in part due to the increase in the total number of papers included in the corpus from those years. That said it is interesting to see that the other five patterns did not see similar increases. As mentioned in Section~\ref{sec:corpus-characteristics}, this may simply be the result of the manner in which the corpus was generated. Nevertheless, this does point towards some patterns being more relevant to so-called modern-day situated visualization than others. Panels are more likely to be used to display 2D visualizations and user interfaces (e.g.,~\cite{fleckRagRugToolkitSituated2022,prouzeauCorsicanTwinAuthoring2020,whitlockHydrogenARInteractiveDataDriven2020,satriadiTangibleGlobesData2022}), ghosts benefit from 3D models being easier to create and more readily accessible (e.g.,~\cite{chenMARVisTAuthoringGlyphBased2020,assorExploringAugmentedReality2023}), and trajectories being used more in analytical contexts of spatio-temporal data (e.g.,~\cite{luoPearlPhysicalEnvironment2023,hubenschmidReLiveBridgingInSitu2022,buschelMIRIAMixedReality2021}).

\section{Common pattern combinations}
\begin{figure*}[t]
\centering
 \includegraphics[width=0.95\linewidth]{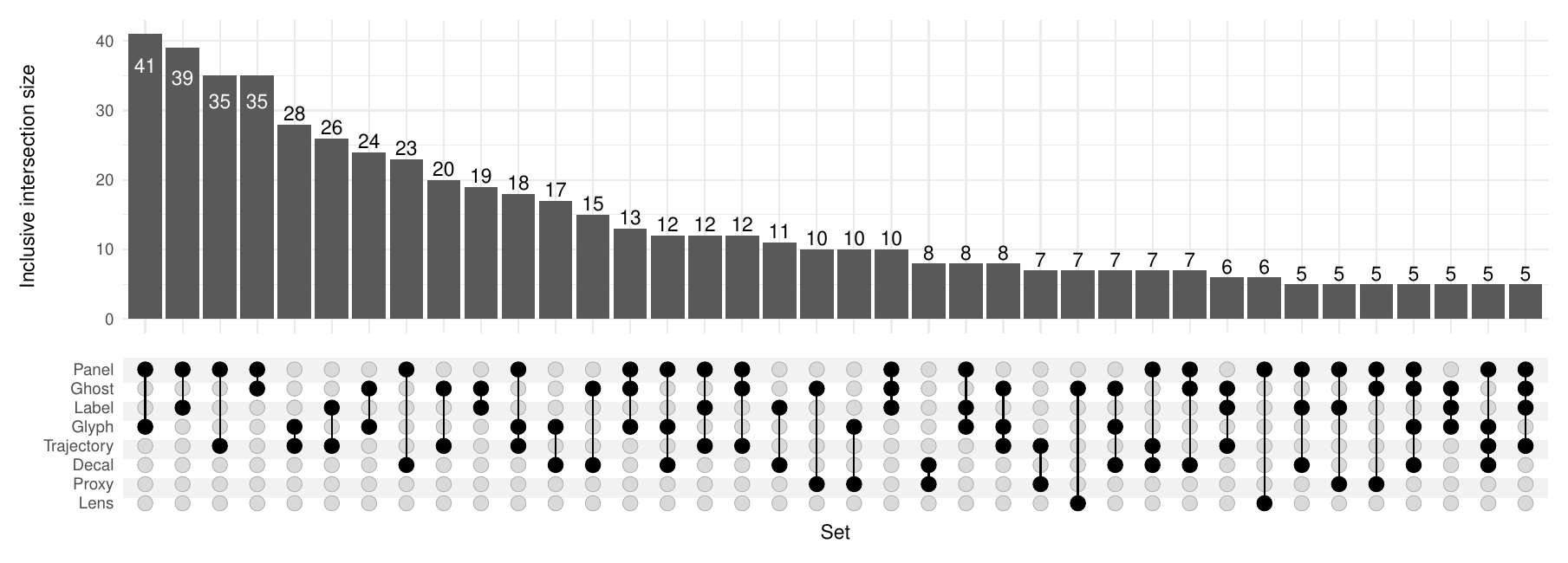}
 \vspace{-4mm}
\caption{An UpSet plot~\cite{lexUpSetVisualizationIntersecting2014,conwayUpSetRPackageVisualization2017} showing the frequency of set combinations between the 10 patterns with the intersect mode (i.e., a set can overlap with other sets). Only sets with a degree of two or higher, and with five or more members are shown.}
\label{fig:survey-upset}
\end{figure*}

Figure~\ref{fig:survey-upset} shows an UpSet plot \cite{lexUpSetVisualizationIntersecting2014,conwayUpSetRPackageVisualization2017} of the frequency of different combinations of the 10 patterns used in our corpus. Note however that, as mentioned in the original manuscript, multiple patterns were assigned to a paper regardless of whether or not the patterns were used simultaneously. For instance, a paper that would demonstrate the use of panels in one prototype and glyphs in another prototype was coded identically to one that used both together.

The 1st to 4th most popular sets are all a combination of panel with another pattern (glyph, label, trajectory, ghost), with panel + decal coming at 8th. This is unsurprising given that these are the most popular patterns (Figure~\ref{fig:survey-pattern-counts}). Even so, this demonstrates the raw versatility that the panel pattern has---even if boring from a novelty perspective. Whenever some information needs to be displayed, be it text, a data visualization, or even a user interface, a panel can be used.

The most popular set which does not include a panel is glyph + trajectory. As mentioned, the two typically go hand. An arrow may point the user to move or look in a specific direction, or be used to highlight some part of the referent which the user should perform some task on. Trajectories can also easily connect multiple glyphs together, akin to a network (e.g., \cite{ensIvyExploringSpatially2017,buschelAugmentedRealityGraph2019}) or a chain of waypoints to follow (e.g., \cite{zollmannFlyARAugmentedReality2014,reitmayrCollaborativeAugmentedReality2004,zhengLocationBasedAugmentedReality2019}).

Trajectory + label is the next most popular non-panel set. This is mainly the use of labels combined with leader lines (e.g.,~\cite{grassetImagedrivenViewManagement2012,tatzgernAdaptiveInformationDensity2016,rekimotoMatrixRealtimeObject1998,guareseAugmentedSituatedVisualization2020,raduWhatCanWe2019,mohrAdaptiveUserPerspective2017,madsenTemporalCoherenceStrategies2016,liShvilCollaborativeAugmented2014}), which count as trajectories connecting the label to the referent. Otherwise, the use of this set is merely coincidental, with the system offering trajectory and label views separately from each other (e.g.,~\cite{fleckRagRugToolkitSituated2022,langnerMARVISCombiningMobile2021,guareseAugmentedSituatedVisualization2020}).

The first set utilizing three patterns is panel + glyph + trajectory. This is effectively a combination of the previous two sets (glyph + trajectory, trajectory + label). A clear use of this set is for navigation, whereby glyphs are used to represent key points of interest, labels provide further information about these points, and trajectories show a guide path between each point (e.g., \cite{liShvilCollaborativeAugmented2014,reitmayrCollaborativeAugmentedReality2004}).

None of the sets in Figure~\ref{fig:survey-upset} include the mirror nor morph patterns. This is to be expected given their already low total occurrences in the corpus. The only noteworthy combinations here involve mirror, which has been combined with glyph, ghost, and panel (i.e.,~\cite{riglingYourFaceVisualizing2023,skreinigARHeroGenerating2022,yuPerspectiveMattersDesign2020,mohrRetargetingVideoTutorials2017,yamaguchiVideoAnnotatedAugmentedReality2020,andersonYouMoveEnhancingMovement2013}). In these instances, the mirror mostly acts to facilitate the AR experience without the use of head-mounted display, and thus can theoretically be combined with any other pattern.

\newpage
\section{Survey conclusions and limitations}
This survey provides further insight as to how the patterns have been used by the papers contained within our corpus. It is not indicative of the entire fields of AR and (situated) visualization, and thus we do not discuss nor provide recommendations for research directions. What we can conclude is that our patterns are expressive enough to cover a wide range of different applications, designs, and use cases. We hope this survey would serve as a starting point for both researchers and situated visualization designers to classify existing work and express new ideas.

\newpage

\renewcommand\tabularxcolumn[1]{m{#1}}
\newcolumntype{C}{>{\centering\arraybackslash}X}
\renewcommand\arraystretch{3}

\begin{table*}[h]
\centering
\begin{tabularx}{\textwidth}{|l|>{\hsize=1.5cm}C|X|l|}
\hline
\textbf{Pattern} & \textbf{Image} & \textbf{References} & \textbf{Count} \\ \hline
Glyph            & \includegraphics[height=1cm]{figures/glyph.pdf}               & \cite{riglingYourFaceVisualizing2023,luoPearlPhysicalEnvironment2023,stanescuModelFreeAuthoringDemonstration2022,hubenschmidReLiveBridgingInSitu2022,zhuMechARspaceAuthoringSystem2022,skreinigARHeroGenerating2022,qianScalARAuthoringSemantically2022,satriadiTangibleGlobesData2022,heyenAugmentedRealityVisualization2022,kayaDynamicContentGeneration2021,langnerMARVISCombiningMobile2021,reipschlagerPersonalAugmentedReality2021,ensUpliftTangibleImmersive2021,hertelAugmentedRealityMaritime2021,buschelMIRIAMixedReality2021,dennlerAugmentedRealityNavigated2021,kongTutorialLensAuthoringInteractive2021,caoExploratoryStudyAugmented2020,chenMARVisTAuthoringGlyphBased2020,kringsDevelopmentFrameworkContextaware2020,yuPerspectiveMattersDesign2020,tainakaGuidelineToolDesigning2020,marquardtComparingNonVisualVisual2020,mohrMixedRealityLight2020,leeEnhancingFirstPersonView2020,luGlanceableAREvaluating2020,herbertPerceptionsIntegratingAugmented2020,langeHiveFiveImmersionPreserving2020,cruzAugmentedRealityApplication2019,buschelAugmentedRealityGraph2019,caoGhostARTimespaceEditor2019,saquibInteractiveBodyDrivenGraphics2019,zhengLocationBasedAugmentedReality2019,werrlichComparingHMDBasedPaperBased2018,blattgersteInSituInstructionsExceed2018,dillmanVisualInteractionCue2018,willettEmbeddedDataRepresentations2017,rambachPOSTERAugmentedThings2017,ensIvyExploringSpatially2017,mohrRetargetingVideoTutorials2017,kimVisARBringingInteractivity2017,datcuHandheldAugmentedReality2016,elsayedSituatedAnalyticsDemonstrating2016,nuernbergerInterpreting2DGesture2016,wuAugmentedRealityInstruction2016,zollmannFlyARAugmentedReality2014,tatzgernTransitionalAugmentedReality2014,kasaharaJackInIntegratingFirstperson2014,zhuAuthorableContextawareAugmented2013,langlotzAudioStickiesVisuallyguided2013,houUsingAnimatedAugmented2013,liuEvaluatingBenefitsRealtime2012,mulloniEnhancingHandheldNavigation2011,pustkaAutomaticConfigurationPervasive2011,mulloniHandheldAugmentedReality2011,mulloniUserExperiencesAugmented2011,zhangContextawareFrameworkSupporting2010,shinguCameraPoseNavigation2010,veneziaContextAwarenessAims2010,chintamaniImprovedTelemanipulatorNavigation2010,wooCAMARContextawareMobile2009,sandorEgocentricSpacedistortingVisualizations2009,mulloniMobilitySocialInteraction2008,nawabJoystickMappedAugmented2007,kobayashiOverlayWhatHumanoid2007,schmalstiegExperiencesHandheldAugmented2007,guvenVisualizingNavigatingComplex2006,reitingerSpatialMeasurementsMedical2005,schmalstiegHandheldAugmentedReality2005,sandorRapidPrototypingSoftware2005,reitmayrCollaborativeAugmentedReality2004,guvenAuthoring3DHypermedia2003,reitmayrLocationBasedApplications2003,reitmayrDataManagementStrategies2003,ulbrichtTangibleAugmentedReality2003,kalkuschStructuredVisualMarkers2002,schmalstiegStudierstubeAugmentedReality2002}                    & 77               \\ \hline
Ghost            & \includegraphics[height=1cm]{figures/ghost.pdf}               & \cite{assorExploringAugmentedReality2023,riglingYourFaceVisualizing2023,luoPearlPhysicalEnvironment2023,zogopoulosAuthoringToolAutomatic2022,liInteractiveAugmentedReality2022,fleckRagRugToolkitSituated2022,stanescuModelFreeAuthoringDemonstration2022,hubenschmidReLiveBridgingInSitu2022,krugCleARSightExploring2022,skreinigARHeroGenerating2022,shinUserOrientedApproachSpaceAdaptive2021,huangAdapTutARAdaptiveTutoring2021,pereiraARENAAugmentedReality2021,sureshkumarAugmentedRealityInternet2021,kayaDynamicContentGeneration2021,somanathHDREnvironmentMap2021,langnerMARVISCombiningMobile2021,reipschlagerPersonalAugmentedReality2021,unluPAIRPhoneAugmented2021,chidambaramProcessARAugmentedRealitybased2021,leivaRapidoPrototypingInteractive2021,ensUpliftTangibleImmersive2021,caoExploratoryStudyAugmented2020,whitlockAuthARConcurrentAuthoring2020,reipschlagerAugmentedDisplaysSeamlessly2020,chenMARVisTAuthoringGlyphBased2020,zhuBISHAREExploringBidirectional2020,kringsDevelopmentFrameworkContextaware2020,whitlockHydrogenARInteractiveDataDriven2020,zhouFineGrainedVisualRecognition2020,yuPerspectiveMattersDesign2020,taharaRetargetableARContextaware2020,yamaguchiVideoAnnotatedAugmentedReality2020,tomkinsBridgingAnalogDigitalDivide2020,mohrTrackCapEnablingSmartphones2019,reipschlagerDesignARImmersive3DModeling2019,caoGhostARTimespaceEditor2019,mahmoodImprovingInformationSharing2019,raduWhatCanWe2019,caoRaInSituVisual2019,werrlichComparingHMDBasedPaperBased2018,herrImmersiveModularFactory2018,blattgersteInSituInstructionsExceed2018,speicherXDARChallengesOpportunities2018,rambachPOSTERAugmentedThings2017,alamAugmentedVirtualReality2017,grubertPervasiveAugmentedReality2017,wangComprehensiveSurveyAugmented2016,hettiarachchiAnnexingRealityEnabling2016,lochComparingVideoAugmented2016,duEdgeSnappingBasedDepth2016,elsayedSituatedAnalyticsDemonstrating2016,mohrRetargetingTechnicalDocumentation2015,speiginerEvolutionArgonWeb2015,kymalainenCoDesigningNovelInterior2014,speiginerEtherealToolkitSpatially2014,galFLAREFastLayout2014,tatzgernTransitionalAugmentedReality2014,kalkofenAdaptiveGhostedViews2013,houUsingAnimatedAugmented2013,iglesiasAttitudeBasedReasoningStrategy2012,yiiDistributedVisualProcessing2012,gervautzAnywhereInterfacesUsing2012,pustkaAutomaticConfigurationPervasive2011,hendersonExploringBenefitsAugmented2011,gruberColorHarmonizationAugmented2010,golparvar-fardApplicationD4AR4Dimensional2009,kalkofenComprehensibleVisualizationAugmented2009,barakonyiAugmentedRealityAgents2008,schoenfelderAugmentedRealityIndustrial2008,junghannsEmployingLocationawareHandheld2008,ledermannAbstractionImplementationStrategies2007,schmalstiegExperiencesHandheldAugmented2007,schallUrban3DModels2007,wagnerHandheldAugmentedReality2006,schmalstiegHandheldAugmentedReality2005,sandorRapidPrototypingSoftware2005,barakonyiAugmentedRealityAgents2005,barakonyiMonkeyBridgeAutonomousAgents2005,wagnerMassivelyMultiuserAugmented2005,barakonyiAgentsThatTalk2004,barakonyiARPuppetAnimated2004,wagnerFirstStepsHandheld2003,tangComparativeEffectivenessAugmented2003,matsushitaIDCAMSmart2003,grassetInteractiveMediatedReality2003,bimberVirtualShowcaseNew2003,macintyreAugmentedRealityNew2001,butzEnvelopingUsersComputers1999,rekimotoMatrixRealtimeObject1998}                    & 90               \\ \hline
Trajectory       & \includegraphics[height=1cm]{figures/trajectory.pdf}               & \cite{luoPearlPhysicalEnvironment2023,fleckRagRugToolkitSituated2022,hubenschmidReLiveBridgingInSitu2022,krugCleARSightExploring2022,zhuMechARspaceAuthoringSystem2022,qianScalARAuthoringSemantically2022,satriadiTangibleGlobesData2022,langnerMARVISCombiningMobile2021,reipschlagerPersonalAugmentedReality2021,leivaRapidoPrototypingInteractive2021,guareseAugmentedSituatedVisualization2021,seigerHoloFlowsModellingProcesses2021,buschelMIRIAMixedReality2021,hubenschmidSTREAMExploringCombination2021,caoExploratoryStudyAugmented2020,guareseAugmentedSituatedVisualization2020,whitlockAuthARConcurrentAuthoring2020,chenMARVisTAuthoringGlyphBased2020,whitlockHydrogenARInteractiveDataDriven2020,yuPerspectiveMattersDesign2020,tainakaGuidelineToolDesigning2020,leivaProntoRapidAugmented2020,herbertUsabilityConsiderationsHand2020,mohrMixedRealityLight2020,buschelAugmentedRealityGraph2019,mohrTrackCapEnablingSmartphones2019,raduWhatCanWe2019,saquibInteractiveBodyDrivenGraphics2019,zhengLocationBasedAugmentedReality2019,gasquesPintARSketchingSpatial2019,caoRaInSituVisual2019,dillmanVisualInteractionCue2018,mohrAdaptiveUserPerspective2017,grubertPervasiveAugmentedReality2017,ensIvyExploringSpatially2017,gruenefeldVisualizingOutofviewObjects2017,tatzgernAdaptiveInformationDensity2016,wuAugmentedRealityInstruction2016,madsenTemporalCoherenceStrategies2016,zollmannFlyARAugmentedReality2014,gauglitzTouchRemoteWorld2014,venta-olkkonenInvestigatingBalanceVirtuality2014,kasaharaJackInIntegratingFirstperson2014,liShvilCollaborativeAugmented2014,kimComparingPointingDrawing2013,grassetImagedrivenViewManagement2012,gervautzAnywhereInterfacesUsing2012,mulloniHandheldAugmentedReality2011,witherIndirectAugmentedReality2011,shinCAMARFutureDirection2009,schmalstiegMobilePhonesPlatform2008,castleVideorateLocalizationMultiple2008,kobayashiOverlayWhatHumanoid2007,leeViewpointStabilizationLive2006,newmanWideAreaTracking2006,reitmayrSemanticWorldModels2005,swanSurveyUserBasedExperimentation2005,schmalstiegHandheldAugmentedReality2005,reitmayrCollaborativeAugmentedReality2004,reitmayrDataManagementStrategies2003,ulbrichtTangibleAugmentedReality2003,wagnerFirstStepsHandheld2003,umlaufARLibAugmentedLibrary2002,schmalstiegStudierstubeAugmentedReality2002,newmanAugmentedRealityWide2001,rekimotoMatrixRealtimeObject1998,fuhrmannCollaborativeAugmentedReality1997}                    & 67               \\ \hline
Decal            & \includegraphics[height=1cm]{figures/decal.pdf}               & \cite{riglingYourFaceVisualizing2023,luoPearlPhysicalEnvironment2023,joshinikhitaDesignFrameworkContextual2022,zogopoulosAuthoringToolAutomatic2022,qianScalARAuthoringSemantically2022,heyenAugmentedRealityVisualization2022,sureshkumarAugmentedRealityInternet2021,reipschlagerPersonalAugmentedReality2021,ensUpliftTangibleImmersive2021,hubenschmidSTREAMExploringCombination2021,heinrichComparisonAugmentedReality2020,tomkinsBridgingAnalogDigitalDivide2020,ferdousWhatHappeningThat2019,strannerHighPrecisionLocalizationDevice2019,subramonyamAffinityLensDataAssisted2019,alvesComparingSpatialMobile2019,zhengLocationBasedAugmentedReality2019,caoRaInSituVisual2019,blattgersteInSituInstructionsExceed2018,lindlbauerRemixedRealityManipulating2018,willettEmbeddedDataRepresentations2017,rambachPOSTERAugmentedThings2017,alamAugmentedVirtualReality2017,joARIoTScalableAugmented2016,elsayedSituatedAnalyticsDemonstrating2016,langlotzNextGenerationAugmentedReality2014,langlotzAudioStickiesVisuallyguided2013,kasaharaExTouchSpatiallyawareEmbodied2013,veasMobileAugmentedReality2013,liuEvaluatingBenefitsRealtime2012,veasExtendedOverviewTechniques2012,zollmannInteractive4DOverview2012,hoangAugmentedViewportAction2010,gruberCitySightsDesign2010,schinkeVisualizationOffscreenObjects2010,wooCAMARContextawareMobile2009,mulloniMobilitySocialInteraction2008,chastineStudiesEffectivenessVirtual2008,schoenfelderAugmentedRealityIndustrial2008,schallHandheldGeospatialAugmented2007,sareikaUrbanSketcherMixed2007,ledermannAPRILHighlevelFramework2005,kingARVinoOutdoorAugmented2005,bonanniAttentionbasedDesignAugmented2005,barakonyiAugmentedRealityAgents2005,reitmayrCollaborativeAugmentedReality2004,reitmayrLocationBasedApplications2003,grassetInteractiveMediatedReality2003,mogilevARPadInterface2002,dorfmuller-ulhaasFingerTrackingInteraction2001}                    & 50               \\ \hline
Morph            & \includegraphics[height=1cm]{figures/morph.pdf}               & \cite{moriSurveyDiminishedReality2017,tatzgernDynamicCompactVisualizations2013,veasExtendedOverviewTechniques2012,sandorEgocentricSpacedistortingVisualizations2009,kalkofenExplosionDiagramsAugmented2009}                    & 5               \\ \hline
Magic Lens       & \includegraphics[height=1cm]{figures/lens.pdf}               & \cite{krugCleARSightExploring2022,reipschlagerPersonalAugmentedReality2021,dennlerAugmentedRealityNavigated2021,lilijaAugmentedRealityViews2019,gsaxnerMarkerlessImagetoFaceRegistration2019,borkBenefitsAugmentedReality2019,eratDroneAugmentedHumanVision2018,dillmanVisualInteractionCue2018,rameauRealTimeAugmentedReality2016,mohrRetargetingTechnicalDocumentation2015,ridelRevealingFlashlightInteractive2014,kalkofenAdaptiveGhostedViews2013,schallGlobalPoseEstimation2009,schallHandheldAugmentedReality2009,mendezImportanceMasksRevealing2009,kalkofenComprehensibleVisualizationAugmented2009,schall3DTrackingUnknown2008,samsetAugmentedRealitySurgical2008,junghannsEmployingLocationawareHandheld2008,schallVirtualRedliningCivil2008,looser3DFlexibleTangible2007,mendezAdaptiveAugmentedReality2007,schallHandheldGeospatialAugmented2007,kalkofenInteractiveFocusContext2007,schallUrban3DModels2007,mendezInteractiveContextdrivenVisualization2006,changIntuitiveIntraoperativeUltrasound2005,swanSurveyUserBasedExperimentation2005,schmalstiegStudierstubeAugmentedReality2002,schmalstiegSewingWorldsTogether1999,viega3DMagicLenses1996,feinerCutawaysGhostingSatisfying1992}                    & 32               \\ \hline
Label            & \includegraphics[height=1cm]{figures/label.pdf}               & \cite{fleckRagRugToolkitSituated2022,zhengSTAREAugmentedReality2022,sureshkumarAugmentedRealityInternet2021,kayaDynamicContentGeneration2021,langnerMARVISCombiningMobile2021,pfeufferARtentionDesignSpace2021,chidambaramProcessARAugmentedRealitybased2021,seigerHoloFlowsModellingProcesses2021,koppelContextResponsiveLabelingAugmented2021,guareseAugmentedSituatedVisualization2020,whitlockAuthARConcurrentAuthoring2020,pohlBodyLayARsToolkit2020,whitlockHydrogenARInteractiveDataDriven2020,wangCAPturARAugmentedReality2020,herbertUsabilityConsiderationsHand2020,prouzeauCorsicanTwinAuthoring2020,xieInteractiveMultiUser3D2020,villanuevaMetaARAppAuthoringPlatform2020,subramonyamAffinityLensDataAssisted2019,raduWhatCanWe2019,luIoTenhancedBidirectionallyInteractive2015,caggianeseSituatedVisualizationAugmented2019,bezerraSmAR2tModelsRuntime2019,herbertGeneralizedRapidAuthoring2018,werrlichComparingHMDBasedPaperBased2018,herrImmersiveModularFactory2018,huoScenariotSpatiallyMapping2018,baumeisterCognitiveCostUsing2017,bachDrawingARCANVASDesigning2017,willettEmbeddedDataRepresentations2017,mohrAdaptiveUserPerspective2017,gruenefeldVisualizingOutofviewObjects2017,kimVisARBringingInteractivity2017,elsayedSituatedAnalyticsDemonstrating2016,tatzgernAdaptiveInformationDensity2016,madsenTemporalCoherenceStrategies2016,orloskyHaloContentContextaware2015,speiginerEvolutionArgonWeb2015,grubertMultiFiMultiFidelity2015,kishishitaAnalysingEffectsWide2014,leppanenAugmentedRealityWeb2014,parkerDataVisualisationTrends2014,speiginerEtherealToolkitSpatially2014,gauglitzTouchRemoteWorld2014,venta-olkkonenInvestigatingBalanceVirtuality2014,kasaharaJackInIntegratingFirstperson2014,langlotzNextGenerationAugmentedReality2014,liShvilCollaborativeAugmented2014,tatzgernHedgehogLabelingView2014,hervasAchievingAdaptiveAugmented2013,zhuAuthorableContextawareAugmented2013,tatzgernDynamicCompactVisualizations2013,kasaharaExTouchSpatiallyawareEmbodied2013,orloskyManagementManipulationText2013,veasMobileAugmentedReality2013,kimComparingPointingDrawing2013,langlotzOnlineCreationPanoramic2012,grassetImagedrivenViewManagement2012,hendersonAugmentedRealityPsychomotor2011,chenMobileAugmentedReality2011,morrisonMobileAugmentedReality2011,pustkaAutomaticConfigurationPervasive2011,hendersonExploringBenefitsAugmented2011,gruberCitySightsDesign2010,mulloniZoomingInterfacesAugmented2010,schmalstiegMobilePhonesPlatform2008,chastineStudiesEffectivenessVirtual2008,castleVideorateLocalizationMultiple2008,leeUnifiedRemoteConsole2007,newmanWideAreaTracking2006,guvenVisualizingNavigatingComplex2006,urataniStudyDepthVisualization2005,schmalstiegAugmentedRealityTechniques2005,reitingerSpatialMeasurementsMedical2005,reitmayrSemanticWorldModels2005,swanSurveyUserBasedExperimentation2005,barakonyiAugmentedRealityAgents2005,reitmayrCollaborativeAugmentedReality2004,tangComparativeEffectivenessAugmented2003,youFusionVisionGyro2001,bellViewManagementVirtual2001,newmanAugmentedRealityWide2001,butzEnvelopingUsersComputers1999,szalavariStudierstubeEnvironmentCollaboration1998,rekimotoMatrixRealtimeObject1998}                    & 85               \\ \hline
Mirror           & \includegraphics[height=1cm]{figures/mirror.pdf}               & \cite{riglingYourFaceVisualizing2023,skreinigARHeroGenerating2022,yuPerspectiveMattersDesign2020,yamaguchiVideoAnnotatedAugmentedReality2020,borkBenefitsAugmentedReality2019,mohrRetargetingVideoTutorials2017,andersonYouMoveEnhancingMovement2013,benkoMirageTableFreehandInteraction2012,bichlmeierVirtualMirrorNew2009,reitmayrWearable3DAugmented2001,bimberAugmentedRealityBackProjection2000}                    & 11               \\ \hline
Proxy            & \includegraphics[height=1cm]{figures/proxy.pdf}               & \cite{satriadiTangibleGlobesData2022,reipschlagerAugmentedDisplaysSeamlessly2020,tomkinsBridgingAnalogDigitalDivide2020,prouzeauCorsicanTwinAuthoring2020,cruzAugmentedRealityApplication2019,reipschlagerDesignARImmersive3DModeling2019,zhengLocationBasedAugmentedReality2019,luIoTenhancedBidirectionallyInteractive2015,werrlichComparingHMDBasedPaperBased2018,lochComparingVideoAugmented2016,tatzgernExploringRealWorld2015,speiginerEvolutionArgonWeb2015,joIoTARPervasive2019,tatzgernDynamicCompactVisualizations2013,tatzgernExploringDistantObjects2013,houUsingAnimatedAugmented2013,veasExtendedOverviewTechniques2012,mulloniHandheldAugmentedReality2011,witherIndirectAugmentedReality2011,hoangAugmentedViewportAction2010,gruberCitySightsDesign2010,markov-vetter3DAugmentedReality2009,ledermannAbstractionImplementationStrategies2007,kalkofenIntegratedMedicalWorkflow2006,newmanWideAreaTracking2006,guvenVisualizingNavigatingComplex2006,ledermannAPRILHighlevelFramework2005,barakonyiAugmentedRealityAgents2005,reitmayrLocationBasedApplications2003,wagnerFirstStepsHandheld2003,kalkuschStructuredVisualMarkers2002,schmalstiegStudierstubeAugmentedReality2002}                    & 32               \\ \hline
Panel            & \includegraphics[height=1cm]{figures/panel.pdf}               & \cite{riglingYourFaceVisualizing2023,luoPearlPhysicalEnvironment2023,joshinikhitaDesignFrameworkContextual2022,zogopoulosAuthoringToolAutomatic2022,fleckRagRugToolkitSituated2022,stanescuModelFreeAuthoringDemonstration2022,hubenschmidReLiveBridgingInSitu2022,krugCleARSightExploring2022,zhuMechARspaceAuthoringSystem2022,skreinigARHeroGenerating2022,qianScalARAuthoringSemantically2022,satriadiTangibleGlobesData2022,tongExploringInteractionsPrinted2022,moriExploringPseudoWeightAugmented2022,zhengSTAREAugmentedReality2022,davariValidatingBenefitsGlanceable2022,huangAdapTutARAdaptiveTutoring2021,pereiraARENAAugmentedReality2021,langnerMARVISCombiningMobile2021,pfeufferARtentionDesignSpace2021,reipschlagerPersonalAugmentedReality2021,chidambaramProcessARAugmentedRealitybased2021,leivaRapidoPrototypingInteractive2021,ensUpliftTangibleImmersive2021,seigerHoloFlowsModellingProcesses2021,buschelMIRIAMixedReality2021,hubenschmidSTREAMExploringCombination2021,luExplorationTechniquesRapid2021,luoInvestigatingDocumentLayout2021,satkowskiInvestigatingImpactRealWorld2021,reichherzerSecondSightFrameworkCrossDevice2021,satkowskiInsituAuthoringAR2021,flickTradeoffsAugmentedReality2021,kongTutorialLensAuthoringInteractive2021,buttnerAugmentedRealityTraining2020,guareseAugmentedSituatedVisualization2020,whitlockAuthARConcurrentAuthoring2020,reipschlagerAugmentedDisplaysSeamlessly2020,zhuBISHAREExploringBidirectional2020,kringsDevelopmentFrameworkContextaware2020,pohlBodyLayARsToolkit2020,whitlockHydrogenARInteractiveDataDriven2020,wangCAPturARAugmentedReality2020,tainakaGuidelineToolDesigning2020,leivaProntoRapidAugmented2020,yamaguchiVideoAnnotatedAugmentedReality2020,herbertUsabilityConsiderationsHand2020,mohrMixedRealityLight2020,prouzeauCorsicanTwinAuthoring2020,blanco-novoaCreatingInternetAugmented2020,cordeilEmbodiedAxesTangible2020,leeEnhancingFirstPersonView2020,xieInteractiveMultiUser3D2020,villanuevaMetaARAppAuthoringPlatform2020,whitlockMRCATSituPrototyping2020,wangUnderstandingAugmentedReality2020,subramonyamAffinityLensDataAssisted2019,whiteAugmentedRealityIoT2019,zachariahBrowsingWebThings2019,reipschlagerDesignARImmersive3DModeling2019,mahmoodImprovingInformationSharing2019,raduWhatCanWe2019,saquibInteractiveBodyDrivenGraphics2019,zhengLocationBasedAugmentedReality2019,caoRaInSituVisual2019,buschelInvestigatingSmartphonebasedPan2019,luIoTenhancedBidirectionallyInteractive2015,bezerraSmAR2tModelsRuntime2019,lagesWalkingAdaptiveAugmented2019,werrlichComparingHMDBasedPaperBased2018,dillmanVisualInteractionCue2018,leeProjectiveWindowsBringing2018,huoScenariotSpatiallyMapping2018,bachHologramMyHand2018,wiesner3DFRCDepictionFuture2017,willettEmbeddedDataRepresentations2017,grubertPervasiveAugmentedReality2017,kimOntologybasedMobileAugmented2017,kimVisARBringingInteractivity2017,joARIoTScalableAugmented2016,datcuHandheldAugmentedReality2016,elsayedSituatedAnalyticsDemonstrating2016,wuAugmentedRealityInstruction2016,radkowskiAugmentedRealityBasedManual2015,speiginerEvolutionArgonWeb2015,grubertMultiFiMultiFidelity2015,ghouaielAdaptiveAugmentedReality2014,pokricAugmentedRealityBased2014,parkerDataVisualisationTrends2014,venta-olkkonenInvestigatingBalanceVirtuality2014,langlotzNextGenerationAugmentedReality2014,ensPersonalCockpitSpatial2014,zhuAuthorableContextawareAugmented2013,langlotzAudioStickiesVisuallyguided2013,houUsingAnimatedAugmented2013,andersonYouMoveEnhancingMovement2013,veasExtendedOverviewTechniques2012,hendersonAugmentedRealityPsychomotor2011,hendersonExploringBenefitsAugmented2011,mulloniHandheldAugmentedReality2011,zhangContextawareFrameworkSupporting2010,hoangAugmentedViewportAction2010,veneziaContextAwarenessAims2010,chintamaniImprovedTelemanipulatorNavigation2010,schinkeVisualizationOffscreenObjects2010,wooCAMARContextawareMobile2009,schallHandheldAugmentedReality2009,shinCAMARFutureDirection2009,goldsmithAugmentedRealityEnvironmental2008,mulloniMobilitySocialInteraction2008,schallVirtualRedliningCivil2008,leeUnifiedRemoteConsole2007,reitingerAugmentedRealityScouting2007,schmalstiegExperiencesHandheldAugmented2007,newmanWideAreaTracking2006,kingARVinoOutdoorAugmented2005,schmalstiegAugmentedRealityTechniques2005,reitingerSpatialMeasurementsMedical2005,reitmayrSemanticWorldModels2005,wagnerMassivelyMultiuserAugmented2005,diverdiLevelDetailInterfaces2004,reitmayrCollaborativeAugmentedReality2004,barakonyiARPuppetAnimated2004,diverdiARWinDesktopAugmented2003,guvenAuthoring3DHypermedia2003,reitmayrDataManagementStrategies2003,matsushitaIDCAMSmart2003,veiglTwohandedDirectInteraction2002,schmalstiegStudierstubeAugmentedReality2002,reitmayrMobileCollaborativeAugmented2001,szalavariStudierstubeEnvironmentCollaboration1998}                    & 131              \\ \hline
\end{tabularx}
\caption{Table of results from our survey ($N$ = 293). A paper may be assigned to more than one pattern.
}
\label{tab:survey-table}
\end{table*}

\clearpage % These four commands fixes the blank space at the start of the reference list
\newpage
\clearpage
\newpage

\bibliographystyle{abbrv-doi-hyperref}

\bibliography{references-survey}